\documentstyle[12pt,preprint,aps,floats,epsfig]{revtex}
\tighten
\def\simlt{\mathrel{\raise.3ex\hbox{$<$\kern-.75em\lower1ex\hbox{$\sim$}}}}
\def\simgt{\mathrel{\raise.3ex\hbox{$>$\kern-.75em\lower1ex\hbox{$\sim$}}}}
\def\ga{\mathrel{\raise.3ex\hbox{$>$\kern-.75em\lower1ex\hbox{$\sim$}}}}
\def\la{\mathrel{\raise.3ex\hbox{$<$\kern-.75em\lower1ex\hbox{$\sim$}}}}

\newcommand{\be}{\begin{equation}}
\newcommand{\ee}{\end{equation}}
\newcommand{\beq}{\begin{equation}}
\newcommand{\eeq}{\end{equation}}
\newcommand{\bea}{\begin{eqnarray}}
\newcommand{\eea}{\end{eqnarray}}
\newcommand{\f}{\frac}

\begin{document}

\preprint{
\noindent
\begin{minipage}[t]{3in}
\begin{flushleft}
July 2001 \\
\end{flushleft}
\end{minipage}
\hfill
\begin{minipage}[t]{3in}
\begin{flushright}
TPI--MINN--01/35\\
UMN--TH--2019/01\\
hep-ph/0107329\\
\vspace*{.7in}
\end{flushright}
\end{minipage}
}

\title{$B\rightarrow X_s \gamma$ in Supersymmetry with
Explicit CP Violation}

\author{D. A. Demir and K. A. Olive
\vspace*{.2in}}
\address{Theoretical Physics Institute, School of Physics and
Astronomy, \\
University of Minnesota, Minneapolis, MN 55455, USA}

\maketitle

\begin{abstract}
We discuss $B\to X_s \gamma$ decay in both constrained and unconstrained
supersymmetric models with explicit CP violation within the minimal flavor
violation scheme by including $\tan\beta$--enhanced large contributions beyond
the leading order. In this analysis, we take into account the relevant
cosmological and collider bounds, as well as electric dipole
moment constraints.  In the unconstrained model, there are portions
of the parameter space yielding a large CP asymmetry at leading
order (LO). In these regions, we find that  the CP phases satisfy certain
sum rules, e.g., the sum of the phases of the $\mu$ parameter and the stop
trilinear  coupling centralize around $\pi$ with a width determined by the
experimental bounds. In addition, at large values of $\tan\beta$, the sign of
the CP asymmetry  tracks the sign of the gluino mass, and the CP
asymmetry is significantly larger than the LO prediction. In the constrained
minimal supersymmetric standard model based on minimal supergravity, we find
that the decay rate is sensitive to the phase of the universal trilinear
coupling.  This sensitivity decreases at large values of the universal gauino
mass. We also show that for a given set of the mass parameters, there exists a
threshold value of the phase of the universal trilinear coupling which grows
with $\tan\beta$ and beyond which  the experimental bounds are satisfied. In
both supersymmetric scenarios, the allowed ranges of the CP phases are wide
enough to have phenomenological consequences.

\vspace*{-.4in}
\end{abstract}

\pacs{PACS numbers: 14.80.Ly, 11.30.Er, 12.60.Jv, 11.30.Pb}

\section{Introduction}

One of the best motivated extensions of the standard model (SM) is softly broken
supersymmetry (SUSY) which provides novel sources for flavor and CP violation
\cite{susy} beyond the Cabibbo--Kobayashi--Maskawa (CKM) picture. While flavor
violation may come from the intergenerational entries of the soft sfermion masses
and tri-linear scalar couplings, their phase content as well as the phases of the
$\mu$ parameter and gaugino masses  provide sources for CP violation. 
Flavor-changing neutral current data puts stringent bounds on flavor
mixings \cite{masiero2}; therefore, such entries must be suppressed as would be the
case if the same unitary rotation which diagonalizes the quark mass matrices also
diagonalizes the squark mass matrices in flavor space. In this minimal flavor
violation (MFV) scheme, which is naturally realized in gravity mediated supersymmetry
breaking and no scale models, flavor violation is minimal as it is generated only by
the CKM matrix. In contrast, the CP  violation is not minimal as it can
spring from both the CKM matrix which leads to flavor--changing processes
($e. g.$ the CP asymmetry in $B\rightarrow X_s \gamma$ decay) and from the
flavor--blind  CP phases of the  soft SUSY--breaking masses with flavor--conserving
processes ($e.g.$ electric dipole  moments (EDMs)) \cite{susy}.  

The effects of the SUSY CP--violation can be manifest in several 
observables such as: the mixing of the Higgs bosons \cite{higgs},
EDMs \cite{edm-ucmssm,oneloop,edm-cmssm,barger,shaaban},
lepton polarization asymmetries in semileptonic decays
\cite{lpol}, the formation of P--wave charmonium and bottomonium
resonances \cite{pmeson}, and CP violation in $B$ meson decays
\cite{ali,ali2,kagan2}  and mixings \cite{indirect}.
Among these observables, the most constraining are the
EDMs \cite{edmexpn,edmexpe,edmexpm} which bound the 
argument of the $\mu$ parameter to be $\simlt \pi/(5 \tan\beta)$, 
leaving the other SUSY phases mostly unconstrained, in both
constrained \cite{edm-cmssm,barger,shaaban} and unconstrained
\cite{edm-ucmssm,oneloop,barger,shaaban} supersymmetry. The constraint on
the phase of the $\mu$ parameter is lifted if the sfermions of the first
two generations weigh $\gg  {\rm TeV}$ as in the effective SUSY scenario
\cite{effsusy} though the EDMs are regenerated at the two--loop level, and can
still  compete with the experimental bounds
\cite{twoloop} in certain portions of the parameter space.

The rare radiative inclusive $B$ meson decay, $B\rightarrow X_s \gamma$ provides a
powerful test of the standard model (SM) and  ``new
physics" such as supersymmetry. The drive to
reduce theoretical uncertainties has led to the computation of next--to--leading order
(NLO) terms in the SM \cite{misiak} and two--doublet models \cite{2hdm}, and
a partial computation to the same order in supersymmetry (SUSY)  \cite{partial,partial1}. The
measurements of the branching ratio at CLEO
\cite{CLEO}, ALEPH \cite{ALEPH} and BELLE
\cite{BELLE} give the combined result\footnote{The combined result is based in part
on an updated value from CLEO of $3.03 \times 10^{-4}$.  We thank G. Ganis and E.
Thorndike for bringing this value to our attention.}
\begin{eqnarray}
\label{exp}
\mbox{BR}\left(B\rightarrow X_s \gamma\right)= \left(3.11 \pm 0.42 \pm
0.21 \right)\times 10^{-4}
\end{eqnarray}
whose agreement with the next--to--leading order (NLO) standard model (SM) prediction \cite{misiak}
\begin{eqnarray}
\label{sm}
\mbox{BR}\left(B\rightarrow X_s \gamma\right)_{\small \mbox{SM}} = \left(3.29 \pm 0.33\right)\times 10^{-4}\:
\end{eqnarray}
is manifest though the inclusion of the nonperturbative effects can modify the result
slightly \cite{kagan1}. That the experimental result (\ref{exp}) and the
SM prediction (\ref{sm}) are in good  agreement shows that the ``new
physics" should lie well above the electroweak scale unless 
certain cancellations occur.   In addition to the
branching ratio, the recent measurement of the CP asymmetry has been
updated to \cite{CLEO001}
\beq
\mbox{A}_{\small \mbox{CP}} \equiv {{\rm BR}\left(b\rightarrow s \gamma\right) -
{\rm BR}\left(\bar b\rightarrow \bar s \gamma\right) \over
{\rm BR}\left(b\rightarrow s \gamma\right) +
{\rm BR}\left(\bar b\rightarrow \bar s \gamma\right)} = (-0.079 \pm 0.108 \pm 0.022)
(1.0 \pm 0.03)
\eeq
implying a 95\% CL range of 
\begin{eqnarray}
\label{expCP}
-0.30\: < \: \mbox{A}_{\small \mbox{CP}}\left(B\rightarrow X_s \gamma\right)\: < \:
0.14
\end{eqnarray}
In the SM, this value is calculated to be rather small: $\mbox{A}_{\small \mbox{CP}}\sim 1\%$
\cite{kagan2}.  

The LEP era has ended with a clear preference to large values of $\tan\beta$
\cite{lep}, for  which it is known that there are $\tan\beta$--enhanced SUSY threshold
corrections \cite{correc} which ($i$) significantly modify \cite{giudice} the leading
order (LO) Wilson coefficients \cite{masiero}, and can ($ii$) dominate the NLO
contributions \cite{giudice,partial}.  In the absence of CP violation, 
weak--scale SUSY satisfies the $B\rightarrow X_s \gamma$
constraints at large
$\tan\beta$ more easily if the
$\mu$ parameter and the trilinear soft masses are of opposite sign
\cite{largetanbet1,largetanbet2}.

In this work we will analyze $B\rightarrow X_s \gamma$ decay in the
minimal supersymmetric extension of the SM in connection with the CP violating phases. 
Above the electroweak breaking scale, the SUSY Lagrangian possesses two global
symmetries: a continious $R$--symmetry, $U(1)_R$, and a Peccei--Quinn symmetry, $U(1)_{PQ}$.
The $U(1)_R$ symmetry is broken by the $\mu$ parameter, the trilinear couplings $A_{f}$ and
the gaugino masses. The $U(1)_{PQ}$ symmetry, however, is sensitive to $\mu$ parameter and Higgs 
bilinear mass parameter $B$ only. Treating the soft masses (and $\mu$ parameter) as spurions it is easy to 
see that theory possesses a full $U(1)_R\times U(1)_{PQ}$ invariance above the 
electroweak scale thereby allowing for the elimination of two dynamical phases.
Though the electroweak breaking leaves only one symmetry to use, the phase
of the Higgs bilinear soft mass $B$ can be always eliminated by rephasing the 
Higgs vacuum expectation values, and therefore, one more phase can be still 
eliminated. Depending on the specific structure of the soft breaking sector this
invariance allows for eliminating one or more phases. In the constrained minimal
supersymmetric standard model (CMSSM) the gaugino masses are universal, and therefore,
all of them can be chosen real leaving the $\mu$ parameter and the trilinear
couplings as the only sources of the supersymmetric  CP violation. In the unconstrained model,
however, one is left with more phases to generate CP violation. In what
follows we will assume a universal phase for the masses of
the $SU(2)$ and  $U(1)_Y$ gauginos, and measure the rest of the phases
(the phase of the $\mu$ parameter, the trilinear couplings, and the gluino mass) with respect to them.
In addition, the sign conventions ($e.g.$ the sign of the $\mu$ parameter 
relative to the trilinear couplings) for soft masses can be fixed from 
the sparticle mass matrices listed in the appendices.

Concerning the calculational precison, we consider the $B\rightarrow X_s \gamma$ decay using 
the LO Wilson coefficients \cite{masiero} and by incorporating beyond leading order (BLO) 
$\tan\beta$--enhanced SUSY threshold corrections \cite{giudice} in the
presence of soft SUSY phases \cite{susy} (which we hereafter call the BLO
scheme to differentiate it from one which includes NLO QCD corrections).
It is important to note that the LO direct CP violation is small 
in  both constrained
\cite{constrainedcpv} and unconstrained
\cite{asymm}  SUSY models once the EDM constraints
\cite{edmexpn,edmexpe,edmexpm} are taken into account
\cite{edm-ucmssm,oneloop,edm-cmssm,barger,shaaban,twoloop}.

Sec. III is devoted to the study of $B\rightarrow X_s \gamma$ decay in an
unconstrained supersymmetric model, i.e., in a model in which soft masses are not
subject to the constraints from supergravity or stringy boundary conditions at
ultra high energies. After determining an appropriate region of the parameter space
that ($i$) suppresses the EDMs and ($ii$) enhances the SUSY threshold corrections,
the allowed ranges of the SUSY phases as well as the resulting CP asymmetry will be
numerically estimated. 
In Sec. IV,  we will perform a detailed analysis of  $b\rightarrow X_s \gamma$ decay
in the constrained MSSM  with explicit CP violation by taking into account the
$\tan\beta$--enhanced SUSY threshold corrections at the weak scale. Since the flavor
mixings in the CMSSM are determined by the CKM matrix, the dominant
contribution to the decay comes from the chargino and charged Higgs exchanges with
flavor--blind phases playing the main role \cite{constrainedcpv}.
Our conclusion are given in section V.
  
\section{General Formalism with CP-Violating Phases}
In this section we study the inclusive radiative $B$--meson decay in a general low
energy SUSY model with nonvanishing soft phases. To a very good approximation,
$B\rightarrow X_s \gamma$ is well approximated \cite{misiak} by the partonic decay
$b\rightarrow s \gamma$ whose analysis can be performed via the effective
hamiltonian 
\begin{eqnarray}
\label{heff}
{\cal{H}}_{eff}= -\f{4 G_F \lambda_{t}}{\sqrt{2}} \sum_{i=1}^{8}
{\cal{C}}_i(Q) {\cal{O}}_i(Q)\:\:,\:\:\left\{ \begin{array}{c}
{\cal{C}}_{1,3,\cdots,6}(Q_W)=0\:, \:
{\cal{C}}_2 (Q_W)=1\:,\\
{\cal{C}}_{7,8}(Q_W)={\cal{C}}_{7,8}^{W} (Q_W) + {\cal{C}}_{7,8}^{H}(Q_W)
+{\cal{C}}_{7,8}^{\chi} (Q_W)\end{array}\right.
\end{eqnarray}
where $\lambda_{t}=K^{*}_{t s} K_{t b}$, $K$ is the CKM matrix, and the
operators ${\cal{O}}_i(Q)$  are defined in \cite{misiak}. Also listed here are the
values of the Wilson coefficients at the weak scale where the electric and
chromoelectric dipole coefficients
${\cal{C}}_{7,8} (Q_W)$ are decomposed in terms of the
$W^{\pm}$, $H^{\pm}$ and $\chi^{\pm}$ contributions as is appropriate in
the MFV scheme.

The negative Higgs searches at LEP prefer those regions of the SUSY parameter space with $\tan\beta \simgt
3.5$ \cite{lep}, and therefore, it is necessary to improve the leading order analysis \cite{masiero,asymm}
by incorporating those SUSY corrections which grow with $\tan\beta$. Indeed, such
non--logarithmic threshold corrections significantly modify tree level Higgs and chargino couplings
\cite{correc,giudice}. The effective lagrangian describing the interactions of quarks with  
$W^{\pm}$, $H^{\pm}$ and the charged Goldstone boson ($G^{\pm}$) is given by
\begin{eqnarray}
\label{lagran}
{\cal{L}}&=&g_{2}\left(Q_W\right)\: \left[K_{t q}\ \overline{t_{L}}\ \gamma^{\mu}\ \ q_{L}\ W_{\mu}^{+}
\: +\: \mbox{h. c.}\right] \nonumber\\
&+&\frac{g_{2}\left(Q_W\right)\ \overline{m_t}\left(Q_W\right)}{\sqrt{2}\ M_W}\:
\left[K_{t s}\ \overline{t_{R}} \left\{ \cot\beta\left(1+ \epsilon_{t s}\ \tan\beta\right)\ H^{+}\: + \:
G^{+} \right\} s_{L}\: +\: \mbox{h. c.}\right]\nonumber\\ 
&+&\frac{g_{2}\left(Q_W\right) \overline{m_b}\left(Q_W\right)}{\sqrt{2}\ M_W
\left(1+\epsilon_{b b}^{\ *}\ \tan\beta\right)}\:
\left[ K_{t b}\ \overline{t_{L}} \left\{ \tan \beta \ H^{+}\: - \:
\left(1+\epsilon_{t b}\ \tan\beta \right)\ G^{+} \right\} b_R \: +\: \mbox{h. c.}\right]
\end{eqnarray} 
where the dimensionless complex coefficients $\epsilon_{b b}$,
$\epsilon_{t s}$ and $\epsilon_{t b}$  represent the SUSY threshold corrections at the associated vertices
\begin{eqnarray}
\label{epsb}
\epsilon_{b b}&=&- \frac{2 \alpha_s}{3
\pi}\: \frac{\mu^{*}}{m_{\widetilde{g}}}\:
{\cal{H}}\left[\frac{M^{2}_{\tilde{b}_1}}{|m_{\widetilde{g}}|^2},
\frac{M^{2}_{\tilde{b}_2}}{|m_{\widetilde{g}}|^2}\right]
-\frac{\alpha_t}{4 \pi}\:\sum_{j=1}^{2}\ \frac{A_t^{*}}{M_{\chi^{\pm}_j}}\: 
\left(C_L\right)_{2 j} \left(C_R^{\dagger}\right)_{2 j}
{\cal{H}}\left[\frac{M^{2}_{\tilde{t}_1}}{M_{\chi^{\pm}_j}^2},
\frac{M^{2}_{\tilde{t}_2}}{M_{\chi^{\pm}_j}^2}\right]\nonumber\\
\epsilon_{t s}&=&\frac{2 \alpha_s}{3
\pi}\: \frac{\mu^{*}}{m_{\widetilde{g}}}\:\sum_{k=1}^{2} \left|C_{\widetilde{t}}^{2
k}\right|^{2}\  {\cal{H}}\left[\frac{M^{2}_{\tilde{t}_k}}{|m_{\widetilde{g}}|^2},
\frac{Q_{12}^{2}}{|m_{\widetilde{g}}|^2}\right]\nonumber\\
\epsilon_{t b}&=&- \frac{2 \alpha_s}{3
\pi} \frac{\mu}{m^{*}_{\widetilde{g}}}\ \sum_{k=1}^{2} \sum_{l=1}^{2}
\left|C_{\widetilde{t}}^{1 k}\right|^{2}\ \left|C_{\widetilde{t}}^{2 l}\right|^{2}\
{\cal{H}}\left[\frac{M^{2}_{\tilde{t}_k}}{|m_{\widetilde{g}}|^2},
\frac{M^{2}_{\tilde{b}_l}}{|m_{\widetilde{g}}|^2}\right]\nonumber\\
&-&\frac{\alpha_t}{4 \pi}\ \sum_{i=1}^{4} \sum_{k=1}^{2} \sum_{l=1}^{2}
\frac{A_t}{M_{\chi^{0}_i}} \left(C_0\right)_{4 i}
\left(C_0^{\dagger}\right)_{3 i}
\left|C_{\widetilde{t}}^{2 k}\right|^{2}\ \left|C_{\widetilde{b}}^{1 l}\right|^{2}\
{\cal{H}}\left[\frac{M^{2}_{\tilde{t}_k}}{M_{\chi^{0}_i}^{2}},
\frac{M^{2}_{\tilde{b}_l}}{M_{\chi^{0}_i}^{2}}\right]
\end{eqnarray}   
where the sfermions of first two generations are assigned an average mass of $Q_{12}$.  
The mixing matrices of charginos $C_{L,R}$, neutralinos $C_{0}$ and squarks
$C_{\widetilde{t},\widetilde{b}}$ are all defined in Appendix A, and the loop function
${\cal{H}}$ is given in Appendix B. In computing (\ref{epsb}), only the gluino  and
Higgsino exchanges are considered as they dominate over those of the  electroweak
gauginos since $\alpha_{2,1}=g_{2,1}^2/(4\pi) \ll \alpha_s=g_s^2/(4\pi)\:,\:
\alpha_t=|h_t|^2/(4\pi)$. A remarkable feature of these vertex corrections is that
they assume nonvanishing values when all SUSY masses are of equal size \cite{correc} 
\begin{eqnarray}
\label{decoup}
\epsilon_{b b}\ \rightarrow \frac{\alpha_s}{3 \pi}\: e^{- i \theta_1}
+\frac{\alpha_t}{8 \pi}\:  e^{- i \theta_2}\ =\
\epsilon_{t b}^{*}\:\: ,\: \: \epsilon_{t s} \rightarrow
-\ \frac{\alpha_s}{3 \pi}\: e^{- i  \theta_1}
\end{eqnarray}
where the two independent phases are given by
$\theta_1=\mbox{Arg}\left[\mu\right]+\mbox{Arg}\mbox{[}m_{\widetilde{g}}\mbox{]}$ and
$\theta_2=\mbox{Arg}\left[\mu\right]+\mbox{Arg}\left[A_t\right]$. Numerically,
$\left|\epsilon_{b b}\right|\sim \left|\epsilon_{t b}\right|\sim \left|\epsilon_{t s}\right|\sim 10^{-2}$,
so that the radiative corrections in (\ref{lagran}) will be ${\cal{O}}(1)$ when $\tan\beta\sim 10^2$.

Using the effective lagrangian (\ref{lagran}) it is straightforward to compute the $W^{\pm}$ and $H^{\pm}$ 
contributions to the electric and chromoelectric dipole coefficients in (\ref{heff})
\begin{eqnarray}
\label{cw}
{\cal{C}}_{7,8}^{W} (Q_W)&=&\frac{3}{2}\
F_{7,8}^{LL}\left[\frac{\overline{m_{t}}^{2}\left(Q_W\right)}{M_W^2}\right]+
\frac{\left[\epsilon_{bb}^{*}-\epsilon_{tb}\right]\
\tan\beta}{1+\epsilon_{bb}^{*}\ \tan\beta} \tilde{F}_{7,8}^{LL}
\left[\frac{\overline{m_{t}}^{2}\left(Q_W\right)}{M_W^2}\right]\\
\label{char}
{\cal{C}}_{7,8}^{H} (Q_W)&=&\frac{1}{2}\cot^2\beta\
F_{7,8}^{LL}\left[\frac{\overline{m_{t}}^{2}\left(Q_W\right)}{M_H^2}\right]+
\frac{1+\epsilon_{ts}^{*}\ \tan\beta}
{1+\epsilon_{bb}^{*}\ \tan\beta}\ \tilde{F}_{7,8}^{LL}
\left[\frac{\overline{m_{t}}^{2}\left(Q_W\right)}{M_H^2}\right]
\end{eqnarray}
where the loop functions are given in Appendix B. These Wilson coefficients now 
possess a  CP--violating character via the $\tan\beta$--enhanced SUSY threshold 
corrections so that they can play a role in the CP asymmetry. Indeed, one may 
treat the BLO piece in ${\cal{C}}_{7,8}^{W} (Q_W)$ as the radiative corection
to the CKM factor $\lambda_t$ (so that the $W$ boson contribution reduces to
its LO form) but this does not eliminate the BLO corrections (and the BLO sources
of CP violation) as the same quantity appears this time in ${\cal{C}}_{7,8}^{H} (Q_W)$
and the chargino contribution.

Even at the LO level, the complex part of the chargino contribution grows linearly
with $\tan\beta$ at large $\tan\beta$ \cite{masiero}. For instance, the 
heavy top squark and charginos give the contribution 
\begin{eqnarray}
\label{wilson}
{\cal{C}}_{7,8}^{\chi} (Q_s)&=&-\sum_{j=1}^{2}\Bigg\{\left| \Gamma_{L}^{1 j} \right|^{2}\
\frac{M_W^2}{M^{2}_{\tilde{t}_1}}\ F_{7,8}^{LL}\left[\frac{M^{2}_{\tilde{t}_1}}{M_{\chi^{\pm}_j}^2}\right]
+ \gamma^{1 j}_{L R}\ \frac{M_W}{M_{\chi^{\pm}_j}}\ F_{7,8}^{LR}
\left[\frac{M^{2}_{\tilde{t}_1}}{M_{\chi^{\pm}_j}^2}\right] \Bigg\}
\end{eqnarray}
where the vertex factors are defined by 
\begin{eqnarray}
\label{vertex}
\Gamma_{L}^{k j}= \left( C_{\widetilde{t}}^{\dagger}\right)^{k 1} \left( C_L\right)^{1 j} -
\frac{\overline{m_{t}}\left(Q_s\right)}{\sqrt{2} M_W \sin \beta}\ \left( C_{\widetilde{t}}^{\dagger}\right)^{k 2}
\left( C_L\right)^{2 j}\:,\:
\Gamma_{R}^{k j}= \frac{ \left( C_{\widetilde{t}}^{\dagger}\right)^{k 1}
\left(C_R\right)^{2 j}}{\sqrt{2}\ \cos\beta \left(1+\epsilon_{b b}^{*}\
\tan\beta\right)}\:,
\end{eqnarray}
with $\gamma^{k j}_{L R}=\left(\Gamma_{L}^{k j}\right)^{*}\ \Gamma_{R}^{k j}$. The dipole
coefficients (\ref{wilson}) are defined at the scale $Q_s$ ($Q_s \gg  Q_W$) which may be 
identified with the masses of the heavy stop or gluino \cite{giudice,partial}.

In the absence of the SUSY CP phases, the theoretical estimate of
$b\rightarrow s \gamma$ is subject to experimental limits on the branching ratio (\ref{exp})
as well as the bounds on the sparticle masses from the direct searches. When the
CP phases are switched on, however, induction of the EDMs is unavoidable and
imposes severe constraints on the parameter space. In supersymmetric models with
explicit  CP violation, the EDM of a fundamental fermion $f$ (first or
second generation leptons or quarks) can receive contributions from
both one-- and two--loop quantum effects:
\begin{eqnarray}
\label{edm}
\frac{{\cal{D}}_f}{e} \: &=& \:\left(\frac{{\cal{D}}_f}{e}\right)_{1-loop}\left[\frac{2 \alpha_s}{3 \pi}\ \frac{m_f\
m_{\widetilde{g}}}{Q^{3}_{12}}\: ,\: \frac{\alpha_2}{4\pi}\ \frac{m_f}{Q^2_{12}} \: ,\: \frac{\alpha_1}{4\pi}\
\frac{m_f}{Q^2_{12}}\right]+\left(\frac{{\cal{D}}_f}{e}\right)_{2-loop}
\end{eqnarray}
where $m_f$ is the mass of the fermion. The arguments of the one--loop contribution corresponds to  the gluino, chargino and neutralino
exchanges, respectively. It is not surprising that this one--loop term
\cite{oneloop} behaves roughly as 
\begin{eqnarray}
\label{edmrel}
\left(\frac{{\cal{D}}_f}{e}\right)_{1-loop}\sim \frac{m_f}{Q^2_{12}}\times \mbox{Im}
\left[\epsilon_{bb}\right]
\left(M_{\tilde{t},\tilde{b}}  \rightarrow Q_{12} \right),
\end{eqnarray}
that is,  the $\tan\beta$--enhanced CP--violating contributions to the Wilson coefficients are directly suppressed by the
one--loop EDMs unless either ($i$) one chooses $Q_{12}$ large enough \cite{oneloop}, or ($ii$) invokes a cancellation mechanism among 
different SUSY contributions. In fact, studies of both unconstrained
\cite{edm-ucmssm,oneloop,barger,shaaban} and constrained
\cite{edm-cmssm,barger,shaaban} supersymmetry show that  the one--loop
EDMs are sufficiently suppressed if $\left|\theta_{\mu}\right| \simlt \pi/(5 \tan\beta)$ with 
no constraints on other  soft phases. In the next section, we will follow the first option, 
that is, we suppress the one--loop EDMs by taking a large enough $Q_{12}$, say, $Q_{12}\simgt 4\ {\rm
TeV}$ \cite{shaaban}.  In Sec. III we will discuss $b\rightarrow s \gamma$ in the
constrained MSSM in which the one--loop EDMs
already agree with the bounds when $\theta_{\mu}$ is close to a
CP--conserving point.
 
The suppression of the one--loop EDMs, however, is not necessarily sufficient, as
there exist two--loop contributions \cite{twoloop} which are 
generated by third generation sfermions. Particularly for down--type fermions,
the two--loop EDMs grow linearly with $\tan\beta$
\begin{eqnarray}
\label{2loop}
\left(\frac{{\cal{D}}_f}{e}\right)_{2-loop}&\sim & |Q_f|\ N_c\ \frac{\alpha\ \alpha_2}{64 \pi^2}\ \frac{m_f m_t}{M_A^2 M_W^2}\ \tan\beta\
\times\nonumber\\
 &&\left[|\mu|\ \sin(2\theta_{\widetilde{t}})\ \sin(\theta_{\mu}+\theta_{A_t})\ f_{\widetilde{t}} - |A_b|\ \sin(2\theta_{\widetilde{b}})\
\sin(\theta_{\mu}+\theta_{A_b})\ f_{\widetilde{b}}\right] 
\end{eqnarray}
where $f=e$ or $d$, and use has been made of the relation $m_b/\cos \beta \sim m_t$. In this expression, the two--loop functions $f_{{\widetilde{b}},
{\widetilde{t}}}$ are defined by 
\begin{eqnarray}
f_{\widetilde{q}}=F\left[\frac{M^{2}_{\tilde{q}_1}}{M_A^2}\right] - F\left[\frac{M^{2}_{\tilde{q}_2}}{M_A^2}\right]
\end{eqnarray}
where the loop function $F\left[x\right]$ is given in Appendix B. This two--loop
contribution is proportional to $\tan\beta/M_A^2$, and can be sizable at large
$\tan\beta$ when the charged Higgs boson is relatively light.   Although one can
partially cancel (\ref{2loop}) by choosing the sbottom sector parameters
appropriately, e.g.,
 $|A_b|\sim |\mu|$, $\theta_{A_t}\sim \theta_{A_b}$ and 
$\theta_{\widetilde{t}}\sim
\theta_{\widetilde{b}}$  at a specific value of $\tan\beta$, in general,
$b\rightarrow s \gamma$ must be analyzed in conjunction with two--loop EDMs in
determining the allowed portions of the SUSY parameter space.

\section{$B\rightarrow X_s \gamma$ in an Effective SUSY Model}
In this section we will discuss the $b\rightarrow s \gamma$ decay in the framework
of an effective SUSY \cite{effsusy} scenario. We choose the first two
generations of sfermions to be heavy enough to suppress their  contributions to EDMs.
Clearly, for $Q_{12}\gg Q_s$ the contributions of the first and second generations
to
$b\rightarrow s \gamma$ are also suppressed \cite{asymm}. 
 
Given the precise fit to the electroweak observables, either both stops
must weigh 
${\cal O}(1)\ {\rm TeV}$ or only a predominantly right--handed stop can be
allowed to weigh as light as 
$\sim Q_W$. In other words, the  stop mixing angle should be small
enough to have 
$\widetilde{t}_2= - \sin \theta_{\widetilde{t}} e^{i
\gamma_{\widetilde{t}}} \widetilde{t}_{L} + \cos \theta_{\widetilde{t}}
\widetilde{t}_{R} \approx \widetilde{t}_{R}$, and $\widetilde{t}_1= \sin
\theta_{\widetilde{t}} e^{-i \gamma_{\widetilde{t}}} \widetilde{t}_{R} +
\cos \theta_{\widetilde{t}}\widetilde{t}_{L} \approx \widetilde{t}_{L}$.
Therefore, a light sparticle spectrum with hierarchy $M_{H},
M_{\tilde{t}_2}, M_{\chi^{\pm}_j}\sim Q_W$, $M_{\tilde{t}_1}\sim
m_{\widetilde{g}}\sim Q_s$ enhances the SUSY contribution to $b\rightarrow
s \gamma$ without spoiling the electroweak precision data
\cite{partial,giudice}. Given that the  SM result (\ref{sm}) is in good
agreement with the experimental result (\ref{exp}) then it is clear that
the contributions of the charged Higgs (which is of the same sign as the
SM result) and chargino--stop loop must largely cancel so  as to agree
with experiment. 

In the limit of degenerate soft masses, the $\tan\beta$--enhanced vertex corrections assume 
nonvanishing values in (\ref{decoup}). In this limiting case, the
$W$--contribution 
${\cal{C}}_{7,8}^{H} (Q_W)$ reduces to its LO form as the radiative
corrections cancel. However, in the very same limit, the charged
Higgs contribution ${\cal{C}}_{7,8}^{H} (Q_W)$ maintains an explicit
dependence on the SUSY phases via $\epsilon_{bb}$ and $\epsilon_{ts}$.
Therefore, unlike the LO  ${\cal{C}}_{7,8}^{H} (Q_W)$, the BLO charged
Higgs contribution obtains a CP--violating character, and together with
the chargino contribution (which contributes to CP violation at
the LO level) they form the two key contributions to the CP asymmetry in
the decay. The vertex correction factors $\epsilon_{bb}, \epsilon_{tb}$
and
$\epsilon_{ts}$ are of ${\cal{O}}(10^{-2})$ in the decoupling limit, and
they induce large corrections at large $\tan\beta$ \cite{giudice}.

Next one observes that in the limiting case of degenerate soft masses (\ref{decoup}), the radiative
corrections to ${\cal{C}}_{7,8}^{W} (Q_W)$ vanish identically in accord with the decoupling 
theorem. However, in the very same limit, the charged Higgs contribution ${\cal{C}}_{7,8}^{H} (Q_W)$
still has an explicit dependence on the SUSY phases via $\epsilon_{bb}$
and
$\epsilon_{ts}$. Therefore, unlike the LO case, the charged
Higgs--induced BLO dipole coefficients acquire a CP--violating potential.
The $\tan\beta$--enhanced vertex corrections $\epsilon_{bb},
\epsilon_{tb}$  and $\epsilon_{ts}$ (at least their dominant pieces 
proportional to $\alpha_s$) assume a value  of $\sim 10^{-2}$ in the
decoupling limit. Actually, they remain close to this value in most of  the
SUSY parameter space. 

Significant stop mass splitting implies that large
logarithms appear when the Wilson
coefficients (\ref{wilson}) are evolved from $Q\sim Q_s$ to $Q\sim Q_W$.
The chargino contribution at the electroweak scale $Q_W$ is
therefore obtained after resummation of such logarithms
\cite{giudice}
\begin{eqnarray}
\label{c7}
{\cal{C}}_{7}^{\chi} (Q_W)&=&\left(\frac{\alpha_s(Q_s)}{\alpha_s(Q_W)}\right)^{\frac{16}{3 \beta_0}}
{\cal{C}}_{7}^{\chi} (Q_s) +\frac{8}{3}\left[ \left(\frac{\alpha_s(Q_s)}{\alpha_s(Q_W)}\right)^{\frac{14}{3
\beta_0}} - \left(\frac{\alpha_s(Q_s)}{\alpha_s(Q_W)}\right)^{\frac{16}{3 \beta_0}}\right]
{\cal{C}}_{8}^{\chi}(Q_s) \nonumber\\
&-&\sum_{j=1}^{2}\Bigg\{ \left(\frac{\alpha_s(Q_s)}{\alpha_s(Q_W)}\right)^{-\frac{4}{\beta_0}}
\left|\Gamma_{L}^{2 j} \right|^{2}\
\frac{M_W^2}{M^{2}_{\tilde{t}_2}}\ F_{7}^{LL}\left[\frac{M^{2}_{\tilde{t}_2}}{M_{\chi^{\pm}_j}^2}\right]
+ {\widetilde{\gamma}}^{2 j}_{L R}\ \frac{M_W}{M_{\chi^{\pm}_j}}\ F_{7}^{LR}
\left[\frac{M^{2}_{\tilde{t}_2}}{M_{\chi^{\pm}_j}^2}\right] \Bigg\}\\
\label{c8}
{\cal{C}}_{8}^{\chi} (Q_W)&=&\left(\frac{\alpha_s(Q_s)}{\alpha_s(Q_W)}\right)^{\frac{14}{3 \beta_0}}
{\cal{C}}_{8}^{\chi} (Q_s)\nonumber\\
&-&\sum_{j=1}^{2}\Bigg\{ \left(\frac{\alpha_s(Q_s)}{\alpha_s(Q_W)}\right)^{-\frac{4}{\beta_0}}
\left|\Gamma_{L}^{2 j} \right|^{2}\
\frac{M_W^2}{M^{2}_{\tilde{t}_2}}\ F_{8}^{LL}\left[\frac{M^{2}_{\tilde{t}_2}}{M_{\chi^{\pm}_j}^2}\right]
+ {\widetilde{\gamma}}^{2 j}_{L R}\ \frac{M_W}{M_{\chi^{\pm}_j}}\ F_{8}^{LR}
\left[\frac{M^{2}_{\tilde{t}_2}}{M_{\chi^{\pm}_j}^2}\right] \Bigg\}
\end{eqnarray}
where the CP--violating parts proportional to $F_{L R}^{7,8}$ are defined via tilded vertex factors 
\begin{eqnarray}
\widetilde{\gamma}^{k j}_{L R}=\gamma^{k j}_{L R}\:+\: \left(\widetilde{\Gamma}_{L}^{k j}\right)^{*}\ \widetilde{\Gamma}_{R}^{k j}\:.
\end{eqnarray}
with  $\widetilde{\Gamma}_{L,R}$ are defined by 
\begin{eqnarray}
\widetilde{\Gamma}_{L}^{k j}= \left( C_{\widetilde{t}}^{\dagger}\right)^{k 1} \left( C_L\right)^{1 j} -
\frac{\overline{m_{t}}\left(Q_W\right)}{\sqrt{2} M_W \sin \beta}\ \left( C_{\widetilde{t}}^{\dagger}\right)^{k
2}\left( C_L\right)^{2 j}\:,\:
\widetilde{\Gamma}_{R}^{k j}= \tan\beta\ \epsilon_{bb}^{*}\ \frac{ \left(
C_{\widetilde{t}}^{\dagger}\right)^{k 1} \left(C_R\right)^{2 j}}{\sqrt{2}\ \cos\beta}\:.
\end{eqnarray}
Obviously, the evolution from $Q_s$ to $Q_W$ depends on the light colored
particle spectrum: when all colored sparticles are heavy $\beta_0=7$,
when the light stop is significantly  lighter than $Q_s$ $\beta_0=41/6$,
and when the light stop, sbottoms and gluino are all much  lighter than
$Q_s$ $\beta_0=9/2$.

After computing the $W$--boson (\ref{cw}), charged Higgs (\ref{char}), and the chargino (\ref{c7},\ref{c8})
contributions to the Wilson coefficients at the weak scale, one can use the standard QCD RGEs for
obtaining the Wilson coefficients at the hadronic mass scale $Q_B\sim m_b$. Then the branching fraction \cite{kagan2}
and the CP asymmetry \cite{kagan1} can be estimated directly. In the calculations below we will take 
$\delta=0.9$, where $\delta$ is a parameter determined by the condition
that the photon energy is above a given threshold
$E_{\gamma} > (1/2) (1-\delta)\ m_{b}$.

The Standard Model CP asymmetry in the $b\rightarrow s \gamma$ decay is
$\sim 1\%$, and therefore, this quantity can be  quite sensitive to
new physics contributions \cite{kagan1}. In an effective supersymmetric
model with LO Wilson coefficients, the CP asymmetry can be as large as
$\sim 8\%$ \cite{asymm} when the charged Higgs, charginos and the lighter
stop all have masses of order the weak scale. In what follows, we will
determine ($i$) the allowed rages of the SUSY CP phases, ($ii$) the size
of the CP asymmetry, and  ($iii$) the correlation between the asymmetry
and the branching ratio. The numerical predictions made via the exact 
expressions of the threshold corrections in (\ref{epsb}) and via the limiting
expressions (\ref{decoup}) are similar to each other. To illustrate the
effects of the threshold corrections on the branching ratio and the
CP asymmetry we fix the values of the parameters as 
$M_{\tilde{t}_2}=M_{H}=250\ {\rm GeV}$, $\left|\mu\right|=\left|A_b\right|=150\
{\rm GeV}$, $\theta_{\widetilde{t}}=\pi/20$, $\theta_{A_b}=\theta_{A_t}$, 
$\tilde{m}_{Q}^{2}=(M_{\widetilde{t}_1})^{2}=(1.2\ {\rm TeV})^2$, and 
$\tilde{m}_{b_R}^{2}=(1\ {\rm TeV})^2$ where the last two masses are
needed to fix the sbottom sector for evaluating the two--loop EDMs (\ref{2loop}).
When evaluating the threshold  correction (\ref{epsb}) we consider their
limiting forms (\ref{decoup}).

\begin{figure}
\hspace*{1in}
\begin{minipage}{7.5in}
\epsfig{file=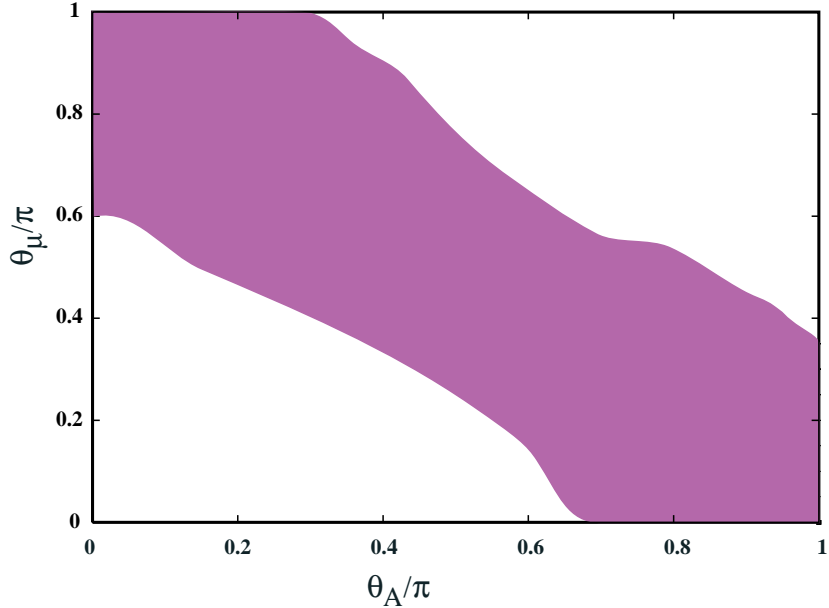,height=3.25in}
\end{minipage}
\vskip .2in
\caption{\label{phases}
{\it The allowed region in the $\theta_{A}$--$\theta_{\mu}$ plane
with LO for $10\leq \tan\beta\leq 60$. The allowed region when BLO Wilson
coefficients  are included is similar.}} 
\end{figure}

Given that the $W$--boson contribution alone is already consistent with the
experiment, and that ${\cal{C}}_{7,8}^{W}(Q_W)$ and ${\cal{C}}_{7,8}^{H} (Q_W)$ have the same sign,
it is clear that some cancellation is needed between the $H^{\pm}$ and  $\chi^{\pm}$
contributions. In the large $\tan\beta$ regime, the difference
between the total SUSY prediction ${\cal{C}}_{7}(Q_W)$ and the
experimental result
${\cal{C}}_{7}^{exp}\sim {\cal{C}}_{7}^{W}(Q_W)$ behaves roughly as
\begin{eqnarray}
\label{rough}
{\cal{C}}_{7}(Q_W)- {\cal{C}}_{7}^{exp}\ \sim\  -\ \left(\cdots\right)\ -\  \tan\beta\
\left(\cdots\right)\ e^{i \theta_2}\: +\: \tan\beta\ \left[\cdots\right]\ \left\{ e^{i \theta_1}\ +\ e^{i\theta_2}\right\}
\end{eqnarray}
where the symbols $\left[\cdots\right]$ and $\left(\cdots\right)$ stand for positive numerical coefficients
with and without the loop suppression, respectively. Shown here is only a rough estimate of the actual coefficients
(\ref{cw}--\ref{c8}) after neglecting several subleading terms. For small or moderate values of $\tan\beta$, 
a cancellation between the chargino and charged Higgs contributions is
needed, and this happens when 
\begin{eqnarray}
\label{phaserel2}
\theta_{\mu}+\theta_{A_t}\leadsto \pi \:.
\end{eqnarray}
At higher values of $\tan\beta$, threshold corrections
become important, and a suppression of such terms occurs when
\begin{eqnarray}
\label{phaserel3}
\theta_{\mu}+\theta_{\widetilde{g}}\leadsto 0
\end{eqnarray}
thus imposing a constraint on the gluino phase. These rough estimates
on the allowed ranges of the  SUSY phases are confirmed in Figure \ref{phases},
which shows the allowed region in $\theta_{A}$--$\theta_{\mu}$ plane 
for the LO approximation. The allowed region when BLO Wilson
coefficients  are included is similar. Here the BLO region is valid so long as the 
gluino mass is positive, $\theta_{\widetilde{g}}=0$. When the sign of the gluino mass is inverted,
$\theta_{\widetilde{g}}=\pi$, the allowed region gets reflected with respect to
the $\theta_{\mu} = \theta_{A}$ line. These direct estimates are in agreement
with the rough estimates in (\ref{phaserel2}) and (\ref{phaserel3})
above.

\begin{figure}
\hspace*{1in}
\begin{minipage}{7.5in}
\epsfig{file=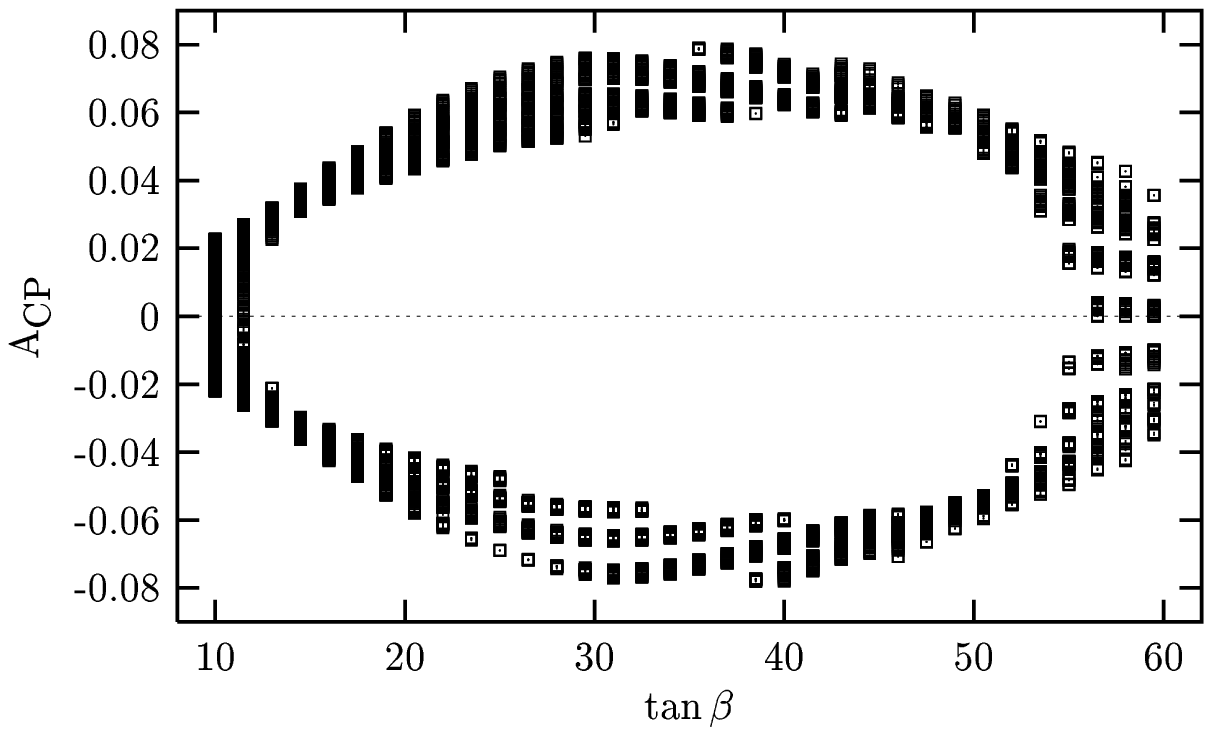,height=3.25in}
\epsfig{file=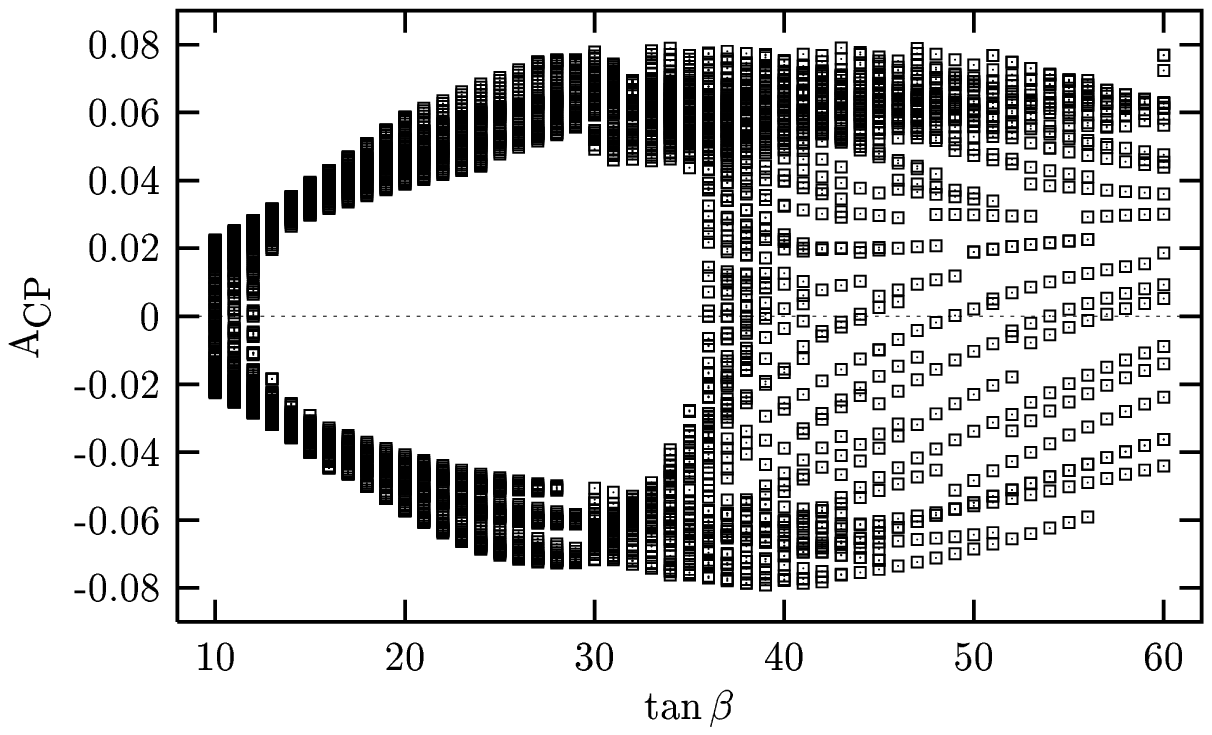,height=3.25in} \hfill
\end{minipage}
\vskip .2in
\caption{\label{asym}
{\it Variation of the CP asymmetry with $\tan\beta$ for LO (upper
window) and BLO (lower window) Wilson coefficients for $10\leq \tan\beta
\leq 60$, and $0\leq (\theta_\mu, \theta_A) \leq \pi$. Individual points correspond to values of the two CP
violating phases taken from Figure \protect\ref{phases}. For $\tan\beta\simlt 35$, 
for which the radiative corrections are small, both asymmetries behave similarly, 
and fall around $2\%$ when  $\tan\beta\sim 10$. The BLO CP asymmetry shown here is 
computed for a positive gluino mass, and to a good approximation, $\mbox{A}_{\small 
\mbox{CP}}\rightarrow - \mbox{A}_{\small \mbox{CP}}$ as $m_{\widetilde{g}}\rightarrow 
-m_{\widetilde{g}}$ at large $\tan\beta$.}}
\end{figure}

In Figure \ref{asym}, we show the variation of the CP asymmetry with
$\tan\beta$ for LO (upper window) and BLO (lower window) Wilson coefficients.
The individual points correspond to a scan over values of $\theta_\mu$ and
$\theta_A$ taken from Figure \ref{phases}. Since the allowed values of the 
phases in the LO and BLO approximations are not the same, the surviving points
in the scans will also differ as seen in comparing the two panels of Figure \ref{asym}.
At LO precision, $\mbox{A}_{\small \mbox{CP}}$ ranges from
$-8\%$ to $8\%$ uniformly, that is, for a given value of $\tan\beta$,
there is no strong preference to positive  or negative values. When BLO
Wilson coefficients are included however, the dependence of 
the CP asymmetry  is modified. We find that there is an observable preference
to  to positive values of  $\mbox{A}_{\small \mbox{CP}}$ distorting the
uniformity of the LO behavior. This particular behaviour of the CP
asymmetry results from our choice of a positive gluino mass, 
$\theta_{\widetilde{g}}=0$. For $\theta_{\widetilde{g}}=\pi$, the graph is 
approximately inverted (relative to that shown in the lower panel of Figure
\ref{asym}), i.e., with a preference to negative values. Therefore, the sign of
the CP asymmetry tracks the phase of the gluino mass for most of the parameter space at
large values of $\tan \beta$.

\begin{figure}
\hspace*{1in}
\begin{minipage}{7.5in}
\epsfig{file=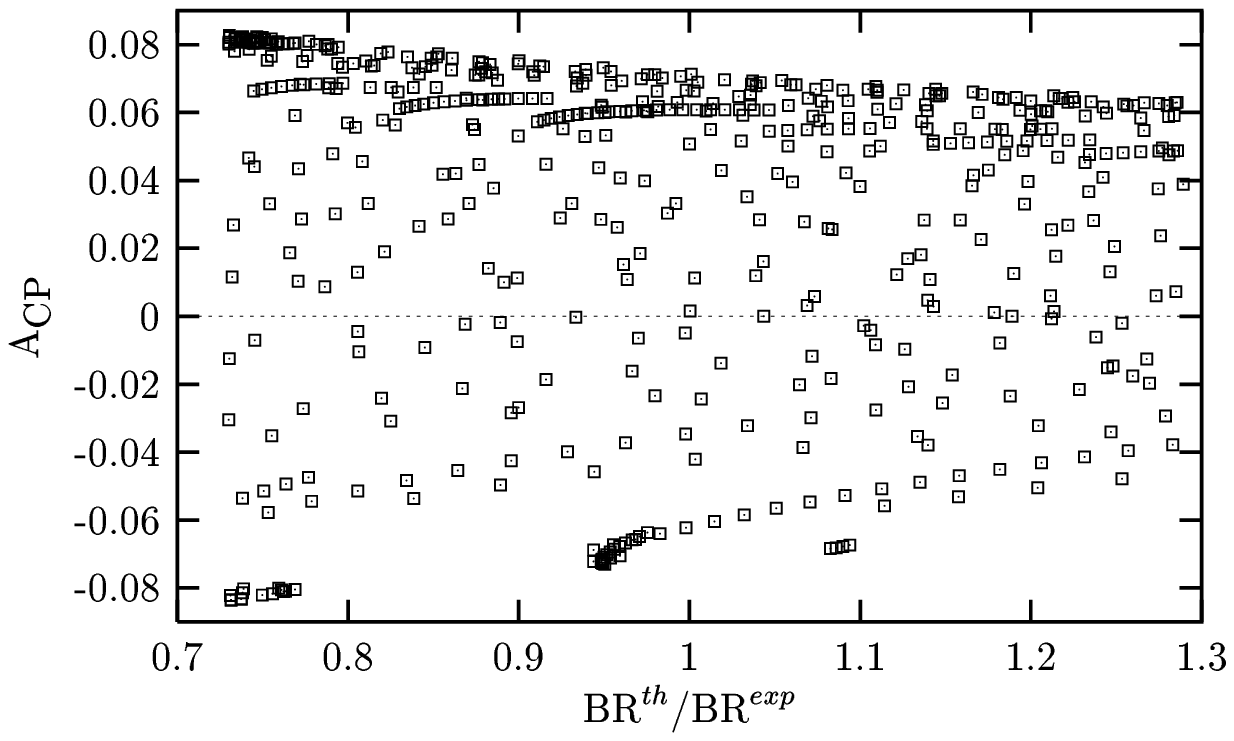,height=3.25in}
\end{minipage}
\vskip .2in
\caption{\label{asym2}
{\it Variation of the CP asymmetry with the branching 
ratio using BLO Wilson coefficients for $m_{\widetilde{g}}>0$.
For most of the points shown, the sign of the CP asymmetry 
is correlated to the sign of the gluino mass.}}
\end{figure}

In general, there are two main reasons that the CP asymmetry is enhanced:
($i$) it is maximized when the branching ratio is minimized, and ($ii$)
it can be maximized  due to specific relations among the Wilson
coefficients depending on the  underlying model. Indeed, it is known that
the asymmetry in the  decay behaves as $\sim 10 \% \times
\left|{\cal{C}}_{8}(m_b)\right|/\left|{\cal{C}}_{7}(m_b)\right|$
\cite{kagan2,wolf}, and reaches the $10\%$ level when the chromoelectric
coefficient has a size similar to the  electric coefficient, as was first
pointed out in the framework of two--doublet models \cite{wolf}. Depicted
in Figure \ref{asym2} is the variation of the CP asymmetry with  the
branching ratio of the decay, where the enhancement of the asymmetry  with
decreasing branching ratio is manifest. In this figure, we assume
$m_{\widetilde{g}} >0$.
For most of the points plotted, there is a correlation  
between the signs of the CP asymmetry and the gluino mass.

In this section we have analyzed the rare radiative $B$ meson decay $B\rightarrow X_s \gamma$ in the 
framework of an effective supersymmetric model \cite{effsusy} in which the
sfermions in first two generations are  heavy enough to suppress the
one--loop EDMs
\cite{edm-ucmssm,oneloop,barger,shaaban}, and the two--loop EDMs
\cite{twoloop} are suppressed by appropriately tuning the stop and
sbottom sector parameters in addition to choosing small stop mixing angles
(which are also required by the electroweak precision
data). 

In this framework we find that, in regions of the SUSY parameter space
where there is a large  CP asymmetry at LO, the BLO effects induce
asymmetries  which are twice as large as the LO asymmetry at sufficiently
large values of
$\tan\beta$. Moreover, certain combinations of the SUSY soft phases
are constrained  to reside  close to CP--conserving points in
order to agree with experimental bounds. The size of the allowed
ranges of the SUSY phases, which is mainly dictated by the experimental 
bounds on the branching fraction and the EDMs, is wide enough to have an
enhanced production of the P--wave bottomonia in lepton colliders
\cite{pmeson}, to have observable CP violation in the Higgs system
\cite{higgs}  in next generation of colliders, and to have observable
leptonic polarization asymmetries in $B$--meson decays \cite{lpol}.
Neither of these phenomena have been observed yet but they are all 
interrelated. First of all, the P--wave  bottomonium  
production rate is determined by $\mbox{Im}[{\cal{C}}_7]$ \cite{pmeson}
which has a richer structure in the BLO case. Secondly, in the 
Higgs system, where the mixing of the opposite CP Higgs bosons depend 
on $\sin(\theta_2)\sin(\theta_{\widetilde{t}})$, there exists sizable
CP violation in the parameter space of Fig. 1 when $\theta_2$ differs
from $\pi$. Moreoever, altough $\theta_{\widetilde{t}}$ is relatively
small (as required by the precision data and EDM constraints) the
CP--violating Higgs mixings are still important with relatively 
light Higgs bosons \cite{higgs} as allowed by $b\rightarrow s \gamma$.
Finally, a simultaneous measurement of the lepton polarization asymmetry and CP asymmetry
in $B\rightarrow K^{\star} \ell^+ \ell^-$ decay provides a good consistencey
check of the 'new physics' contributions where both depend on the 
imaginary parts of the dipole coefficients ${\cal{C}}_{7,8}$ (and other
relevant coefficients), and they are necessarily enhanced in the BLO case \cite{lpol}.

\section{$B\rightarrow X_s \gamma$ in the CMSSM}

In this section, we will analyze the implications of the SUSY CP violating
phases on  $b\rightarrow s \gamma$  in the framework of  the
minimal supergravity model, or equivalently, the constrained MSSM or CMSSM
\cite{CMSSM}.
In the CMSSM, universal gaugino masses $m_{1/2}$,
scalar masses $m_0$ (including those of the Higgs multiplets) and
trilinear supersymmetry breaking parameters $A_0$ are input at the
supersymmetric grand unification scale. In this framework, the Higgs
mixing parameter $\mu$ can be derived (up to its phase which does not run) from the
other MSSM parameters by imposing the electroweak vacuum conditions for any given
value of $\tan \beta$.   In the CMSSM, there are only two physical phases
to consider.
Thus, given the set of input parameters determined
by $\{ m_{1/2}, m_0, |A_0|,\tan\beta,\theta_\mu$ and $\theta_A \}$, the entire
spectrum of sparticles can be derived.

In CMSSM, the low energy mass spectrum is completely controlled by the
GUT--scale parameters above, and one cannot decouple the first two
generations of sfermions (though they are nearly  degenerate to an
excellent approximation); therefore, the Wilson coefficients at the
SUSY--breaking scale (\ref{wilson}) must be updated by taking
$Q_{12}\sim Q_s$. Consequently, the contribution of squarks
in the first two generations give 
\begin{eqnarray}
\label{wilson2}{\cal{C}}_{7,8}^{\chi} (Q_s)&=& \left[
{\cal{C}}_{7,8}^{\chi} (Q_s)\ \mbox{in Eq.}\
(\ref{wilson})\right]\nonumber\\ &+&\sum_{j=1}^{2}\Bigg\{
\left| \widehat{\Gamma}_{L}^{1 j} \right|^{2}\
\frac{M_W^2}{Q_s^2}\ F_{7,8}^{LL}\left[\frac{Q_s^2}{M_{\chi^{\pm}_j}^2}\right]
+\left(\widehat{\Gamma}_{L}^{1 j}\right)^{\star}\ \widehat{\Gamma}_{R}^{1 j}\
\frac{M_W}{M_{\chi^{\pm}_j}}\ F_{7,8}^{LR} \left[\frac{Q_s^2}{M_{\chi^{\pm}_j}^2}\right]\Bigg\}
\end{eqnarray}where $\widehat{\Gamma}_{L,R}$  is obtained by replacing
the stop mixing matrix $C_{\widetilde{t}}$ with the identity matrix in
(\ref{vertex}). If the CMSSM spectrum admits the lighter stop to be as
light as $\sim Q_W$ (which can occur when large values of $A_0$ are
assumed), then the analysis of
$b\rightarrow s
\gamma$ proceeds  as in the last section, in particular, the chargino
contribution at the weak scale is given by  (\ref{c7}) and (\ref{c8})
where ${\cal{C}}_{7,8}^{\chi} (Q_s)$ is now given by (\ref{wilson2}).

The remaining colored sparticles typically have masses
around the SUSY breaking scale, $Q_s$.
For such a hierarchy of the masses, the analysis of the Wilson coefficients differs 
from the previous cases, in particular, the chargino contribution at the SUSY breaking scale is now 
given by 
\begin{eqnarray}
\label{wilson3}
{\cal{C}}_{7,8}^{\chi} (Q_s)&=&\sum_{j=1}^{2}\Bigg\{
\left| \widehat{\Gamma}_{L}^{1 j} \right|^{2}\
\frac{M_W^2}{Q_s^2}\ F_{7,8}^{LL}\left[\frac{Q_s^2}{M_{\chi^{\pm}_j}^2}\right]
+\left(\widehat{\Gamma}_{L}^{1 j}\right)^{\star}\ \widehat{\Gamma}_{R}^{1 j}\
\frac{M_W}{M_{\chi^{\pm}_j}}\ F_{7,8}^{LR}
\left[\frac{Q_s^2}{M_{\chi^{\pm}_j}^2}\right]\nonumber\\
&-&\sum_{k=1}^{2}\left\{ \left| \Gamma_{L}^{k j} \right|^{2}\
\frac{M_W^2}{M^{2}_{\tilde{t}_k}}\ F_{7,8}^{LL}\left[\frac{M^{2}_{\tilde{t}_k}}{M_{\chi^{\pm}_j}^2}\right]
-\gamma^{k j}_{L R}\ \frac{M_W}{M_{\chi^{\pm}_j}}\ F_{7,8}^{LR}
\left[\frac{M^{2}_{\tilde{t}_k}}{M_{\chi^{\pm}_j}^2}\right]\right\} \Bigg\}
\end{eqnarray}   
which can be rescaled to the electroweak scale via QCD running
\begin{eqnarray}
\label{c7p}
{\cal{C}}_{7}^{\chi} (Q_W)&=&\left(\frac{\alpha_s(Q_s)}{\alpha_s(Q_W)}\right)^{\frac{16}{3 \beta_0}}
{\cal{C}}_{7}^{\chi} (Q_s) +\frac{8}{3}\left[ \left(\frac{\alpha_s(Q_s)}{\alpha_s(Q_W)}\right)^{\frac{14}{3
\beta_0}} - \left(\frac{\alpha_s(Q_s)}{\alpha_s(Q_W)}\right)^{\frac{16}{3 \beta_0}}\right]
{\cal{C}}_{8}^{\chi}(Q_s)\\
\label{c8p}
{\cal{C}}_{8}^{\chi} (Q_W)&=&\left(\frac{\alpha_s(Q_s)}{\alpha_s(Q_W)}\right)^{\frac{14}{3 \beta_0}}
{\cal{C}}_{8}^{\chi} (Q_s)
\end{eqnarray}
where $\beta_0=7$, as the colored spectrum below $Q_s$ is just that of the
SM. The formulae (\ref{wilson2},\ref{wilson3})  give possible changes in
the effective theory at the weak scale in analyzing $b\rightarrow s
\gamma$, which  will be numerically analyzed by taking into account
cosmological and collider constraints. 

Due to the restricted parameter set, and the strong correlation among the 
masses of the sparticles, there are a number of experimental constraints which must be
considered.  The most important of these are provided by LEP searches for sparticles
and Higgs bosons~\cite{Junk}, the latter constraining the sparticle spectrum
indirectly via radiative corrections. The kinematic reach for charginos was
$m_{\chi^\pm} = 104$~GeV, and the LEP limit is generally close to this value, within
the CMSSM framework. The LEP searches for sleptons impose $m_{\tilde e} >
97$~GeV, $m_{\tilde \mu} > 94$~GeV and $m_{\tilde \tau} > 80$~GeV for $m_\chi <
80$~GeV.  Other important sparticle
constraints are those on stop squarks ${\tilde t}$: $m_{\tilde t} >
94$~GeV for $m_\chi < 80$~GeV from LEP, and $m_{\tilde t} \ga 115$~GeV
for $m_\chi \la 50$~GeV from the Fermilab Tevatron collider~\cite{stopT}.

The lower limit on the mass of a Standard Model Higgs boson imposed by the
combined LEP experiments
is $113.5$~GeV~\cite{LEPHiggs}. This lower limit also applies to the
MSSM for small $\tan \beta$, even if squark mixing is maximal. In the
CMSSM, maximal mixing is not attained, and the $e^+ e^- \to Z^0 + h$
production rate is very similar to that in the Standard Model~\cite{ZH},
for all values of $\tan \beta$. As is well known, a 2.9-$\sigma$ signal
for a Higgs boson weighing about $115^{+1.3}_{-0.7}$~GeV has been
reported~\cite{LEPHiggs}.  

In addition, the BNL E821 experiment has recently reported \cite{Brown:2001mg} a new
value for the anomalous magnetic moment of the muon: $g_\mu - 2 \equiv 2 \times
a_\mu$,  which yields an apparent discrepancy with the Standard Model prediction
at the level of 2.6 $\sigma$: 
$
\delta a_\mu \; = \; (43 \pm 16) \times 10^{-10}.
$
This result has been well studied in the CMSSM \cite{gm2}, resulting in a strong
preference to $\mu > 0$ and relatively low values of $m_{1/2}$ and $m_0$.

In addition to the phenomenological constraints, one must also carefully consider the
resultant relic density of the LSP \cite{EHNOS}.  This constraint generally disfavors
large values of either of the SUSY breaking mass scales $m_0, m_{1/2}$ except in some
well defined regions of parameter space where an enhanced annihilation cross section
ensures that  the relic density $\Omega h^2 < 0.3$ \cite{efgosi,others}. These 
include regions where co-annihilations are important \cite{cos}, s-channel 
pseudo-scalar exchange is important \cite{efgosi}, or in the focus point region at
large $m_0$ \cite{fp}. The co-annihilation region is important at large $m_{1/2}$ and
the allowed regions due to pseudo-scalar exchange occur at large $m_{1/2}$ and large
$m_0$ at large $\tan \beta$. As we will see, the sensitivity to the phase
$\theta_A$ occurs at relatively low $m_{1/2}$ and $m_0$. This is the region favored by
the recent results from the $g-2$ experiment.

\begin{figure}
\vspace*{-0.75in}
\begin{minipage}{7.5in}
\epsfig{file=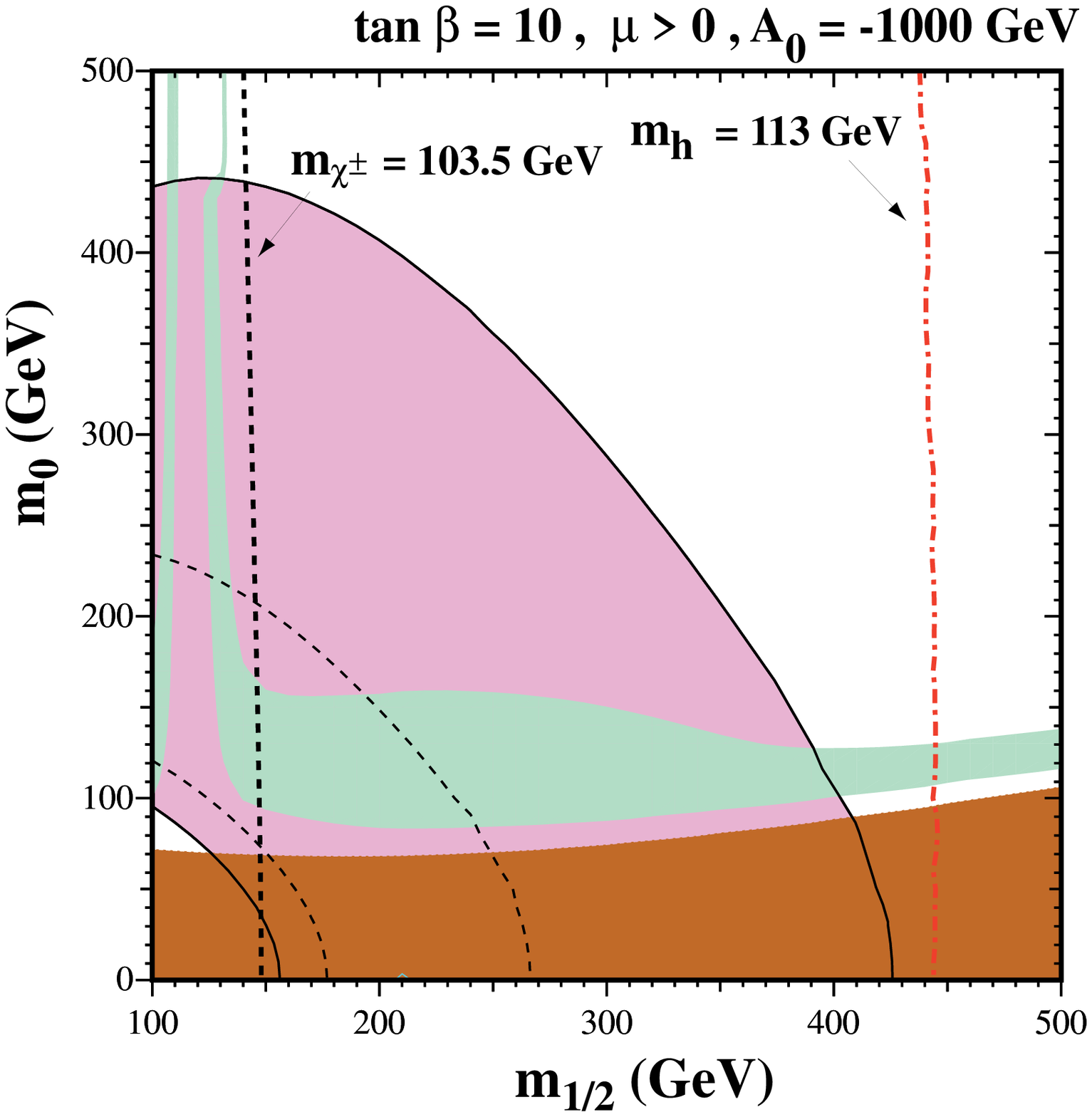,height=3.25in}
\epsfig{file=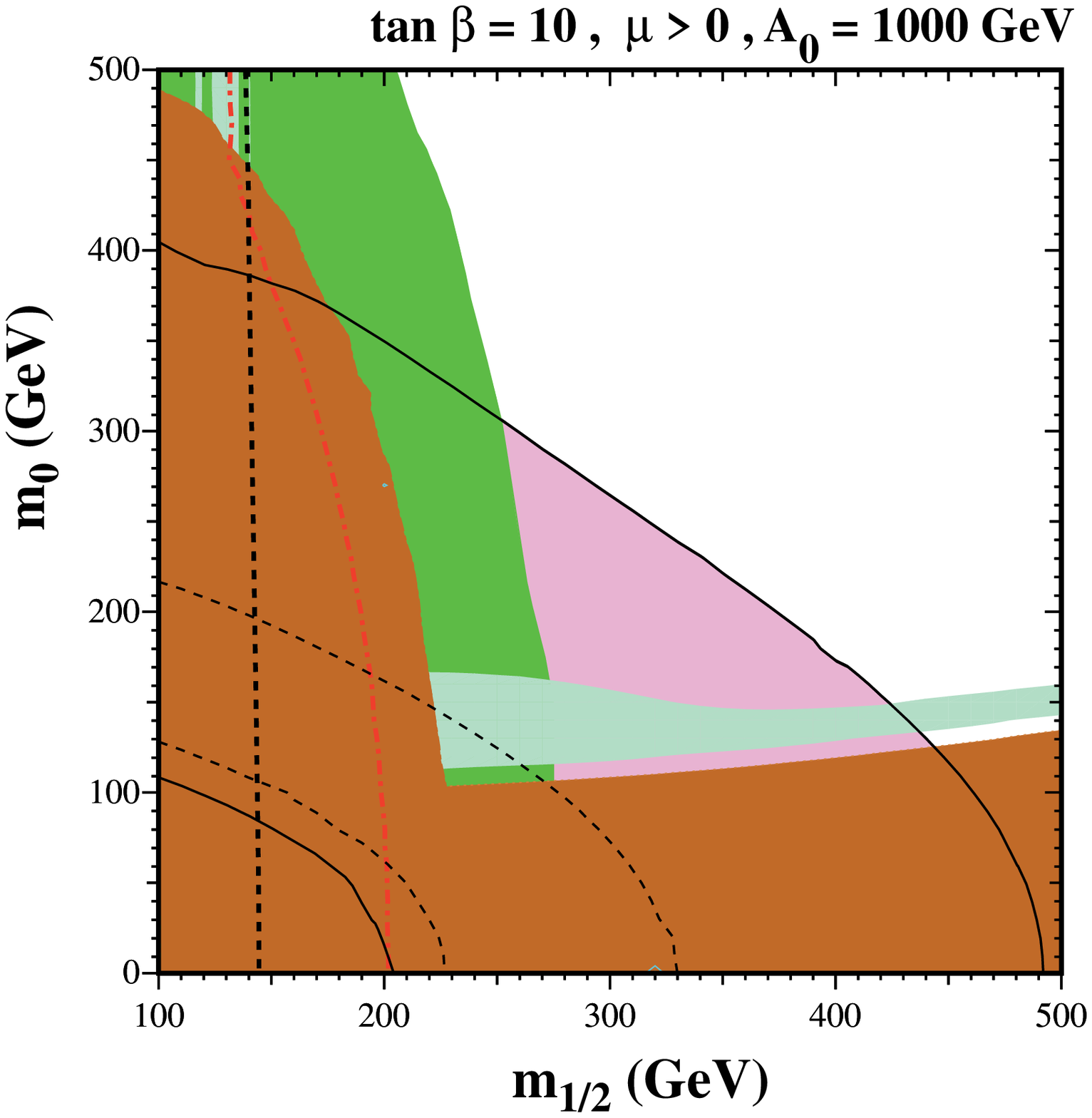,height=3.25in} \hfill
\end{minipage}
\vskip .2in
\caption{\label{plane}
{\it The $(m_{1/2}, m_0)$ planes for $\tan \beta = 10$, $\mu > 0$ and  (a) $A_0 =
-1000$ GeV, (b) $A_0 = 1000$ GeV.  In each case we have assumed $m_t = 175$ and  
$m_b(m_b)^{\overline {MS}}_{SM} = 4.25$~GeV. The near-vertical (red) dot-dashed
lines are the contours
$m_h = 113$~GeV, as evaluated using the {\tt FeynHiggs} code.
The medium (dark green) shaded regions are excluded by $b
\rightarrow s \gamma$.
The light (turquoise) shaded areas are the cosmologically
preferred
regions with \protect\mbox{$0.1\leq \Omega h^2 \leq 0.3$}. In the
dark (brick red) shaded regions, the LSP is either ${\tilde \tau}_1$ or $\tilde t_1$,
so this region is excluded. The regions allowed by the E821 measurement of
$a_\mu$ at the 2-$\sigma$ level are shaded (pink) and bounded by solid
black lines, with dashed lines indicating the 1-$\sigma$ ranges. 
}}
\end{figure}

In Figure \ref{plane}, we show the cosmological and phenomenological
constraints in the
$m_{1/2},m_0$ parameter plane for $\tan \beta = 10$. 
The very dark (red) regions  correspond to either
$m_{\tilde \tau_1} < m_{\chi}$, or $m_{\tilde t_1} < m_{\chi}$ where the subscript 1
denotes the lighter of the 
$\tilde
\tau$ or $\tilde t$ mass eigenstates. These regions 
are ruled out by the requirement that the LSP be neutral.  For $A_0 < 0 $, the
excluded  region is found at low values of $m_0$ where $\tilde \tau_1$ is the LSP. For
$A_0 > 0$, the $\tilde \tau_1$ LSP region is similar, but there is in addition a
region where the stop is lighter than the neutralino at low $m_{1/2}$ and extends to
$m_0 \simeq 500$  GeV (in much of this region the stop is actually tachyonic).  We
show as (red) dash-dotted, nearly vertical lines the
$m_h = 113$~GeV contour calculated using {\tt FeynHiggs}~\cite{Heinemeyer:2000yj}.
For $A_0 < 0$, the limit is quite strong and excludes values of $m_{1/2} \la 440$
GeV.  For $A_0 > 0$, the Higgs mass limit is not significant.  The (dashed) bound on
the chargino mass from LEP excludes very low values of 
$m_{1/2} \la 140$ GeV.
 The branching
ratio for $b \rightarrow s \gamma$
excludes a dark (green) shaded  area at low $m_{1/2}$.
For these parameter choices, this region is only present for $A_0 > 0$.
The (pink) medium-shaded region corresponds to the region {\em preferred}
by the recent $g-2$ experiment. The region bounded by the solid curves
corresponds to the 2-$\sigma$ preferred region, whereas the dashed curves
correspond to 1-$\sigma$. Finally, the (turquoise) light shaded region
corresponds to the parameter space for which the neutralino relic density
falls in the range $0.1 < \Omega h^2 < 0.3$.

The CP violating phases in the CMSSM have been studied
\cite{edm-cmssm,barger,shaaban} in relation to their effects on EDMs. 
Generally, it is found that while the phase of $\mu$ is strongly constrained
by the EDMS to be close to either of its two CP conserving values, the
phase of
$A$ is largely unconstrained. For this reason, in what follows, we will
set $\theta_\mu = 0$ and concentrate on the effect of $\theta_A$ on the
rate for
$b \rightarrow s \gamma$. 

\begin{figure}
\hspace*{1in}
\begin{minipage}{7.5in}
\epsfig{file=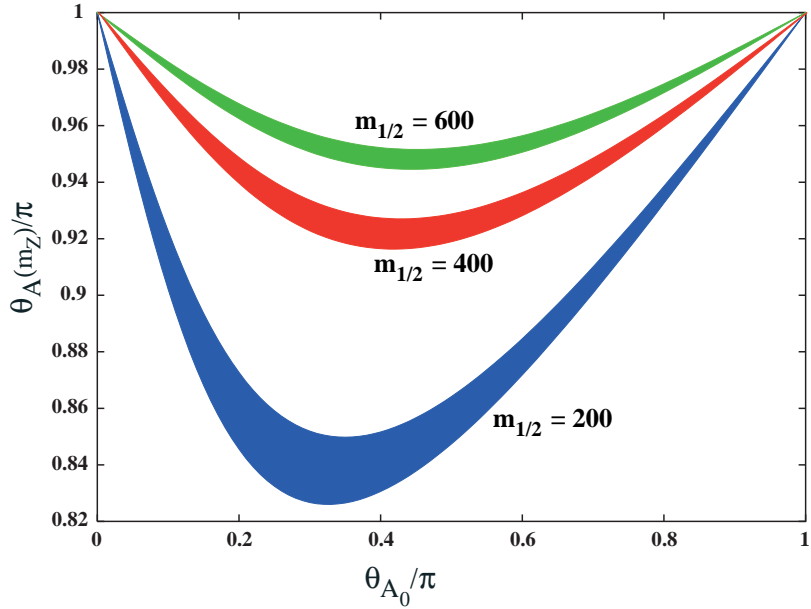,height=3.25in}
\end{minipage}
\vskip .2in
\caption{\label{theta2}
{\it The value of the phase of A at the weak scale as a function of the
input phase, $\theta_{A_0}$ for $m_{1/2}$ = 200, 400, and 600 GeV in the
range $\tan \beta =$ 10 - 30. Here $m_0$ is fixed at 400 GeV and $|A_0| =
1000$ GeV.}}
\end{figure}

In the CMSSM, we specify the phases at the GUT scale.  Since $\theta_\mu$,
is not affected by the running of the renormalization group equations, the value of
$\theta_\mu$ at the weak scale is identical to its input value.  However, the real and
imaginary parts of $A$ run differently and hence, the weak scale phase, $\theta_A$
differs from its input value $\theta_{A_0}$. In Figure \ref{theta2}, we show the
resulting phase of $A_t$ at the weak scale as a function of the input value for
several choices of $m_{1/2} = 200, 400$, and 600 GeV, with $m_0$ = 400 GeV
and $|A_0| = 1000$ GeV. The shaded regions show the dependence with
respect to
$\tan \beta$ between 10 and 30. The values of $\theta_A$ for $\tan \beta = 10$ are
given by the lower edge of each shaded region, whereas the result for $\tan \beta =
30$ is given by the upper edge.  As one can see in the figure, the
tendency of the RGE's is to drive $A_t$ towards real values, particularly
at higher values of $m_{1/2}$. Nevertheless, the resulting phase at the
weak scale is sufficient to have a strong impact on  the branching ratio
of $b
\rightarrow s \gamma$ as we now discuss. 

\begin{figure}
\hspace*{1in}
\begin{minipage}{7.5in}
\epsfig{file=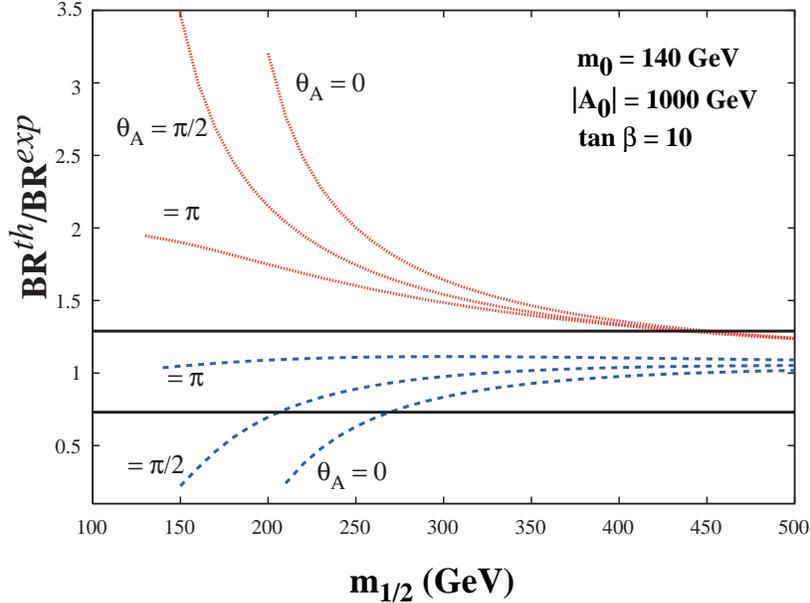,height=3.25in}
\end{minipage}
\vskip .2in
\caption{\label{tb10}
{\it The calculated branching ratio for $b \to s \gamma$ scaled to the experimental
value of 3.11 $\times 10^{-4}$ as a function of $m_{1/2}$ for $\tan \beta = 10, m_0 =
140$ GeV, and $|A_0| = 1000$ GeV. Shown are the branching ratios for $\theta_{A_0} = 0,
\pi/2$, and $\pi$. Curves for $\mu > 0$ are shown as dotted and red, while curves for
$\mu < 0$ are dashed and blue. The horizontal solid black lines delimit the 95 \% CL
experimental range.}}
\end{figure}

\begin{figure}
\hspace*{1in}
\begin{minipage}{7.5in}
\epsfig{file=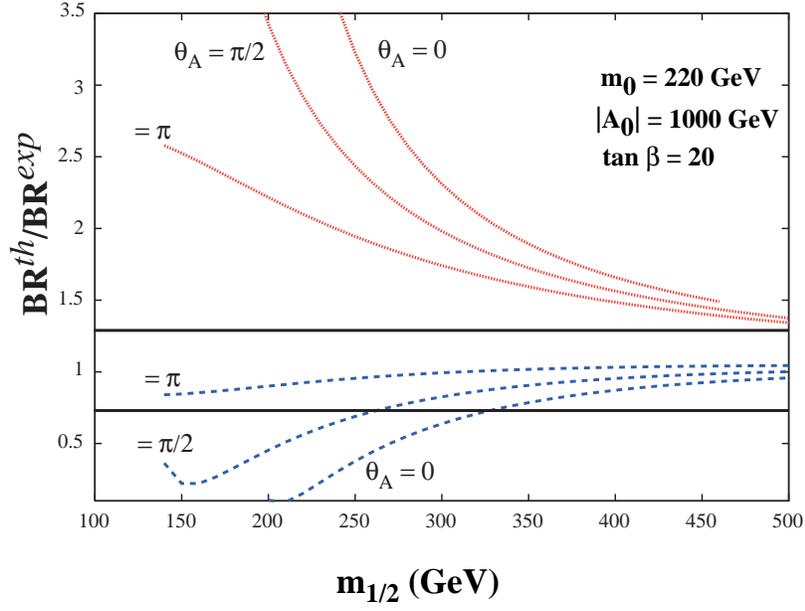,height=3.25in}
\end{minipage}
\vskip .2in
\caption{\label{tb20}
{\it As in Fig. \protect\ref{tb10} for $\tan \beta = 20$ and $m_0 =
220$}}
\end{figure}

\begin{figure}
\hspace*{1in}
\begin{minipage}{7.5in}
\epsfig{file=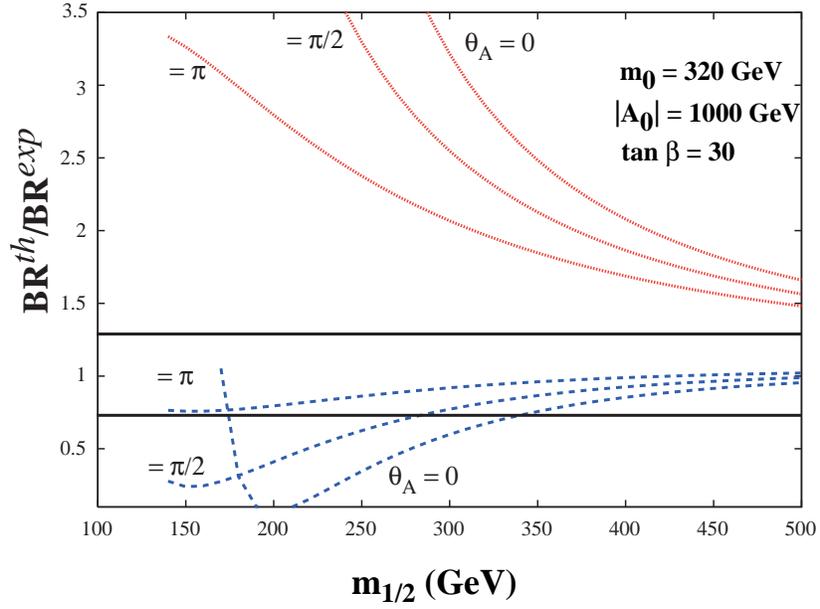,height=3.25in}
\end{minipage}
\vskip .2in
\caption{\label{tb30}
{\it As in Fig. \protect\ref{tb10} for $\tan \beta = 30$ and $m_0 =
320$}}
\end{figure}

In Figures \ref{tb10} -- \ref{tb30}, we show the calculated branching
ratio of
$B
\rightarrow X_s \gamma$ (scaled to the experimental value of $3.11 \times
10^{-4}$) as a function of $m_{1/2}$ for $\theta_{A_0} = 0, \pi/2$, and
$\pi$ for both $\mu < 0$ (dotted curves) and $\mu > 0$ (dashed curves).
The horizontal lines correspond to the 95 \% CL range (2.27 -- 4.01)
$\times 10^{-4}$. We have fixed the value of $m_0 = 140$, for $\tan \beta = 10$, which
is the value which best yields a good relic density (for this value of $|A_0|$).
As one can see the curves calculated for $\mu < 0$ show a branching ratio which is
too large unless $m_{1/2} \ga 430$ GeV.  At low $m_{1/2}$, even though the phase of
$A_0$ has a strong effect on the calculated branching ratio, it can not reduce it to
the experimentally allowed value. On the other hand, for $\mu > 0$, we see that while
all values of $m_{1/2}$ are allowed by $b\rightarrow s \gamma$ for $\theta_{A_0} = \pi$,
there is a strong dependence on $\theta_A$ thus enabling one to set a limit on the
phase of
$A_0$ from $b \rightarrow s \gamma$, at least for some values of
$m_{1/2}$. One should bear in mind that while the range $m_{1/2} < 440$
GeV is nominally excluded by the Higgs mass limit, the theoretical
uncertainty in the calculated Higgs mass is about 3 GeV. The $m_h = 110$
GeV contour would lie at $m_{1/2} = 320$ GeV in Figure \ref{plane}b.

In Figures \ref{tb20} and \ref{tb30}, we show the analogous behavior of
the  branching ratio for $\tan \beta = 20$ and 30 respectively. For the
larger values of 
$\tan \beta$, it is not possible to satisfy the cosmological constraint with a single
value of $m_0$.  However, we have chosen a value of $m_0$ which best fits the
cosmological region for both signs of $\mu$ and $A$. In addition, we note that the
dependence of the branching ratio on $m_0$ is relatively weak. Thus had we
in fact varied $m_0$ with $m_{1/2}$ (to insure a proper relic density),
the curves in Figures \ref{tb20} and \ref{tb30} would differ only very slightly.

The sensitivity of the branching ratio to the phase of $A_0$ is also strong at
the larger values of $\tan \beta$ as seen in Figures \ref{tb20} and \ref{tb30}.
As $\tan \beta$ is increased, one is pushed to higher values of $m_{1/2}$
and therefore we are able to set bounds on $\theta_{A_0}$ for a wider range in
$m_{1/2}$. The Higgs mass bounds are also weaker at higher $\tan \beta$. 

\begin{figure}
\hspace*{1in}
\begin{minipage}{7.5in}
\epsfig{file=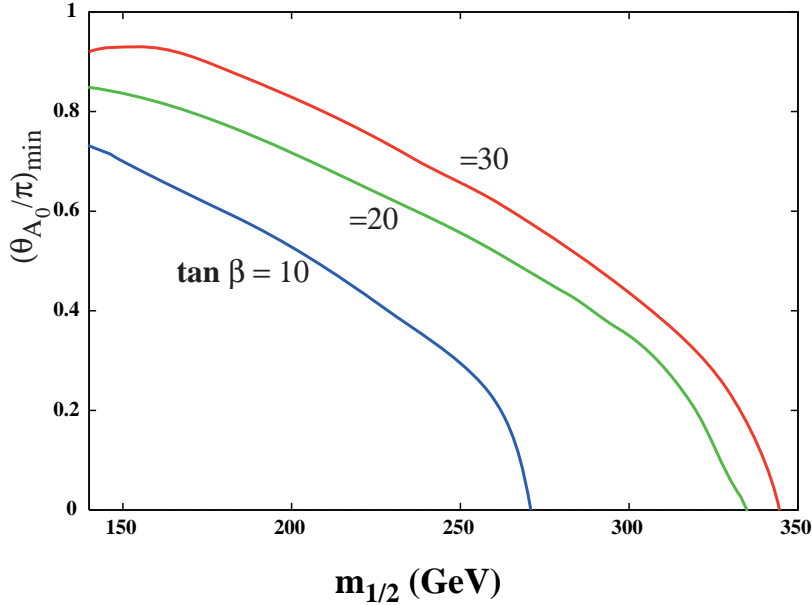,height=3.25in}
\end{minipage}
\vskip .2in
\caption{\label{theta1}
{\it The lower bound on $\theta_{A_0}$ as a function of $m_{1/2}$ for $\tan \beta =
10, 20$ and  30.  Values of $m_0$ are the same as in Figures \protect\ref{tb10} -
\protect\ref{tb30}. Note that at $m_{1/2}\approx 170\ {\rm GeV}$ there will be 
a small window at $\theta_{A_0}\approx 0$ which is allowed for $\tan\beta=30$. We do
not include this curve in the figure. 
}}
\end{figure}

As noted above, for a given range in $m_{1/2}$, we are able to use the constraints
from $b \to s \gamma$ to set a limit on $\theta_{A_0}$.  These limits are summarized
in Figure \ref{theta1}. Shown there is the {\em lower} limit to $\theta_{A_0}$ as a
function of $m_{1/2}$ for the three values of $\tan \beta$ indicated. Values of
$m_0$ were taken from Figures \ref{tb10} - \ref{tb30}. One can also read off from this
figure the minimal value of $m_{1/2}$ such that all values of the phase,
$\theta_{A_0}$, are allowed.

\begin{figure}
\hspace*{1in}
\begin{minipage}{7.5in}
\epsfig{file=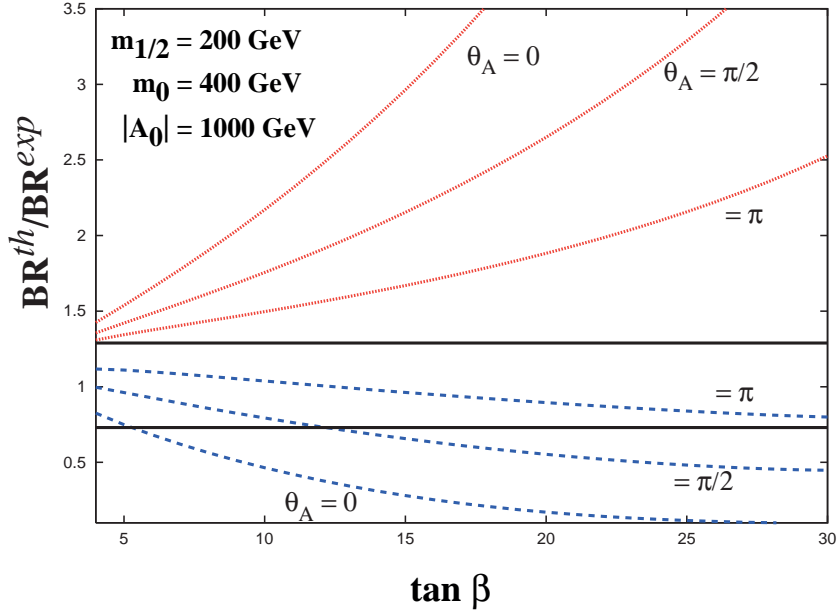,height=3.25in}
\end{minipage}
\vskip .2in
\caption{\label{tb1}
{\it The calculated branching ratio for $b \to s \gamma$ scaled to the experimental
value of 3.11 $\times 10^{-4}$ as a function of $\tan \beta$ for $m_{1/2} = 200$ GeV,
$m_0 = 400$ GeV, and $|A_0| = 1000$ GeV. Shown are the branching ratios for
$\theta_{A_0} = 0, \pi/2$, and $\pi$. Curves for $\mu > 0$ are shown as dotted and red,
while curves for
$\mu < 0$ are dashed and blue. The horizontal solid black lines delimit the 95 \% CL
experimental range.}}
\end{figure}

\begin{figure}
\hspace*{1in}
\begin{minipage}{7.5in}
\epsfig{file=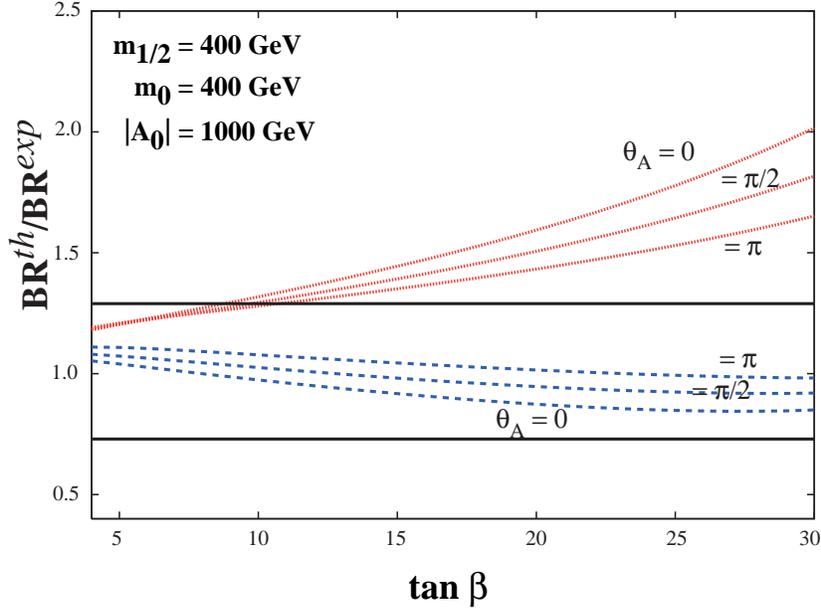,height=3.25in}
\end{minipage}
\vskip .2in
\caption{\label{tb2}
{\it As in Fig. \protect\ref{tb1} for $m_{1/2} = 400$.}}
\end{figure}

Finally, in Figures \ref{tb1} and \ref{tb2}, we show the branching ratio as a function
of  $\tan \beta$ for fixed $m_0 = 400$ GeV , $|A_0|$ = 1000 GeV, and $m_{1/2}$ = 200
and 400 GeV respectively. Recall that the sensitivity of these curves to $m_0$ (at
least for relatively low $m_0$) is rather small. As one can see, at
small values of $m_{1/2}$, there is a strong dependence of the branching
ratio on the phase. At higher values of $m_{1/2}$, not only is the
dependence weaker, but also as one can see, for a given value of $\tan
\beta$, either all phases are allowed or forbidden by the experimental
constraint on $b \to s \gamma$.

\section{Summary}

In this work we have performed a thorough study of the constraints on the
SUSY soft phases from $B\to X_s \gamma$ decay using existing experimental
bounds. Our analysis was restricted to the MFV scheme in which 
flavor violation occurs only through the CKM matrix. We
considered both unconstrained as well as constrained supersymmetric
models.

In particular, we considered an unconstrained supersymmetric model, for
which low energy supersymmetry bears no imprint of the stringy boundary
conditions at ultra high energies. In particular, we chose a particle
spectrum such that the charged Higgs boson, charginos, and the lighter
stop are relatively light (with masses of order the weak scale).  In this
case,  we found that:
($i$) the sum of the phases of the $\mu$ parameter and the stop trilinear 
coupling must take values around the CP--conserving point
$\pi$, with a width determined by experimental uncertainties; ($ii$)
the sum of the phases of the $\mu$ parameter and the gluino mass tends to
zero within the present experimental precision; ($iii$) the LO CP
asymmetry swings from
$-8\%$ to $+8\%$ where positive and negative values are equally possible;
($iv$) the inclusion of the $\tan\beta$--enhanced corrections widens the
allowed range of $\tan\beta$ values, though it does not lead to a
significant change in the size of the CP asymmetry except for the fact
that ($a$) it can be approximately twice as large as the LO prediction,
and ($b$) it tends to follow the sign of the gluino mass for most of the
parameter space, for $\tan\beta\sim {\cal{O}}(50)$. Consequently, there
exist observable effects of the SUSY threshold corrections  at large
$\tan\beta$. If experiments measure a large CP asymmetry
(compared to the SM expectation), this may be interpreted as having
originating from the soft phases and as an indication for weak scale
SUSY. 

After analyzing the implications of the CP violating phases
in an unconstrained supersymmetric model, we turned to a detailed
discussion of the CMSSM with explicit CP violation. In our numerical
estimates, we took the $\mu$ parameter to be real as implied by earlier
studies of the EDM constraints, and we explored the
$b\rightarrow s \gamma$ branching ratio in regions of the parameter space
allowed by the cosmological as well as other collider constraints
displayed in Figure
\ref{plane}. In particular, we discussed the impact of the phase of the
universal trilinear coupling at the GUT scale on the branching ratio. 
As the associated figures suggest, the branching ratio is quite sensitive
to $\theta_{A_0}$, at least for moderately small values of $m_{1/2}$. 
As we have shown, for positive values of $\mu$, while $\theta_{A_0} =
\pi$, generally produced branching ratios in agreement with the
experimental bounds, deviations from this CP conserving point can lead to
branching ratios which are not compatible with experiment.  We have also
seen that the phase of the universal trilinear coupling can not
ameliorate the inconsistency at low $m_{1/2}$ when
$\mu<0$.

\noindent{ {\bf Acknowledgments} } \\
\noindent  
This work was supported in part by DOE grant
DE--FG02--94ER--40823. We would like to thank Toby Falk, Gerri Ganis and
Stefano Rigolin for many helpful conversations.

\section*{Appendix A: Masses and Mixings of SUSY Particles}
\setcounter{equation}{0}\def\theequation{A.\arabic{equation}}
Given that the intergenerational mixings are proportional to the
corresponding quark masses, the weak eigenstate squarks are
approximately the mass eigenstates for the first two generations.
However, the third generation squarks as well as the  gauginos and
Higgsinos can mix strongly after electroweak symmetry breaking. Since
the chargino mass matrix
$M_{\chi^\pm}$ is not hermitian, it is convenient to consider
$\widetilde{M_L}^2=M_{\chi^\pm}^{\dagger}\cdot M_{\chi^\pm}$ and
$\widetilde{M_{R}}^2=M_{\chi^\pm} \cdot M_{\chi^\pm}^{\dagger}$ which are
hermitian and can be diagonalized by
\begin{eqnarray}
\label{sf}
{C}_{\alpha}^{\dagger}\, \widetilde{M_{\alpha}}^2\, {C}_{\alpha} =
\mbox{diag.}\left(M^{2}_{\alpha_1}, M^{2}_{\alpha_2}\right)\:\:\mbox{with}\:\:
{C}_{\alpha}  = \left( \begin{array}{cc} \cos \theta_{\alpha} & - \sin \theta_{\alpha}\, e^{- i
\gamma_{\alpha}}\\
\sin \theta_{\alpha}\, e^{ i \gamma_{\alpha}} & \cos \theta_{\alpha}\end{array}\right) \cdot \left(
\begin{array}{cc} e^{i\eta_{\alpha}} & 0 \\ 0 & e^{i\rho_{\alpha}} \end{array}\right)
\end{eqnarray}
The squark mass matrices
$\widetilde{M_{\widetilde{t, b}}}^2$ are similarly diagonalized. The relative sign between $A_{t, b}$
and $\mu$ is set by the corresponding off--diagonal element of the sfermion mass matrix: $m^{2}_{12}=
-m_t(A_t - \mu \cot \beta)$ for stops, and $m^{2}_{12}=-m_b(A_b - \mu \tan\beta)$ for sbottoms.
In eq. (\ref{sf}), the explicit expressions for the angle parameters are given by
\begin{eqnarray}
&&\gamma_{\widetilde{t}}=-\mbox{Arg}\left[A^*_t - \mu\cot\beta\right]\:\:,\:\:
 \tan\ 2 \theta_{\widetilde{t}}= \frac{ 2 m_t \left|A^*_t -
\mu\cot\beta\right|}{M^2_{\tilde{t}_{L}} -M^2_{\tilde{t}_{R}}-(1/6) \cos 2\beta
\left(5 M_Z^2 - 8 M_W^2\right)}\nonumber\\
&&\gamma_{\widetilde{b}}=-\mbox{Arg}\left[A^*_b - \mu\tan\beta\right]\:\:,\:\:
 \tan\ 2 \theta_{\widetilde{b}}= \frac{ 2 m_b \left|A^*_b -
\mu\tan\beta\right|}{M^2_{\tilde{b}_{L}} -M^2_{\tilde{b}_{R}}+(1/6) \cos 2\beta
\left(M_Z^2 - 4 M_W^2\right)}\nonumber\\ &&\gamma_{L}=-\mbox{Arg}\left[M_2  +\mu
\cot\beta\right]\:\:,\:\:
\tan\ 2 \theta_L= \frac{ \sqrt{8} M_W \sin \beta \left| M_2  +
\mu \cot\beta\right|}{M_2^{2} + |\mu|^{2} + 2 M_W^{2} \cos 2 \beta}\nonumber\\
&&\gamma_{R}=-\mbox{Arg}\left[M_2  + \mu^{*} \tan\beta\right]\:\:,\:\:
\tan\ 2 \theta_R= \frac{ \sqrt{8} M_W \cos
\beta \left| M_2  + \mu^{*} \tan \beta \right|}{M_2^{2} - \left|\mu\right|^{2} - 2
M_W^{2} \cos 2 \beta}
\end{eqnarray}
where $0\leq \theta_{\alpha} \leq \pi/2$ to ensure $M^{2}_{\alpha_1} >
M^{2}_{\alpha_2}$. Finally, one can choose
$\eta_{\widetilde{t},L}=\rho_{\widetilde{t},L}=0$, and
\begin{eqnarray}
\eta_R&=&\mbox{Arg}\left[c_R\left( M_2 c_L +\sqrt{2} M_W \sin\beta s_L e^{i\gamma_L}\right)
+s_R e^{-i \gamma_R}\left(\sqrt{2} M_W \cos\beta c_L + \mu s_L
e^{i\gamma_L}\right)\right]\\
\rho_R&=&\mbox{Arg}\left[c_R\left(-\sqrt{2} M_W \cos\beta  s_L e^{-i \gamma_L}-\mu c_L\right)
-s_R e^{i \gamma_R}\left(-M_2 s_L e^{-i \gamma_L}+ \sqrt{2} M_W \sin\beta c_L \right)\right]
\end{eqnarray}
with $s_{L,R}=\sin \theta_{L,R}$ and $c_{L,R}=\cos \theta_{L,R}$. The mixing matrix $C_{\alpha}$ guarantees
that both chargino masses are real positive with $M^{2}_{\chi_1}> M^{2}_{\chi_2}$ and $M^{2}_{\widetilde{t,b}_1}>
M^{2}_{\widetilde{t,b}_2}$.

Finally, the neutralinos are described by a $4\times4$ mass matrix
\begin{eqnarray} M^{0} \ =\ \left(
\begin{array}{cccc} M_1& 0 & M_Z s_w \cos \beta & - M_Z s_w \sin \beta\\
0 & M_2 & - M_Z c_w \cos \beta & M_Z c_w
\sin \beta\\ M_Z s_w \cos \beta & - M_Z c_w \cos \beta & 0 & -\mu \\ -
M_Z s_w \sin \beta & M_Z c_w \sin \beta &
-\mu & 0\end{array}\right)
\end{eqnarray}
which can be diagonalized numerically via
\begin{eqnarray}
C_{0}^{T} M^{0} C_{0} =\mbox{diag.}\left(M_{\chi^{0}_1}, \cdots,
M_{\chi^{0}_4}\right)
\end{eqnarray}
where $M_{\chi^{0}_1}< \cdots < M_{\chi^{0}_4}$.

\section*{Appendix B: Loop Functions}
\setcounter{equation}{0}\def\theequation{B.\arabic{equation}}
The two--point function entering the expressions of $\epsilon_{bb,ts,tb}$ (\ref{epsb}) is
\begin{eqnarray}
{\cal{H}}\left[x, y\right]&=&\frac{x}{(1-x)\ (x-y)}\ \ln{x} + \frac{y}{(1-y)\ (y-x)}\ \ln{y}
\end{eqnarray}
with ${\cal{H}}\left[1, 1\right]=-1/2$, and ${\cal{H}}\left[1, 0\right]=-1$. 

The loop functions entering the expressions of the Wilson coefficients (\ref{cw}--\ref{wilson})
are given by
\begin{eqnarray}
F_{7}^{LL}\left[x\right]&=&\frac{x\left(7 - 5 x - 8 x^2\right)}{36 (x-1)^3}+\frac{x^2\left(3 x - 2\right)}{6 (x-1)^4}
\ln{x}\nonumber\\
F_{7}^{LR}\left[x\right]&=&\frac{5 - 7 x}{6 (x-1)^2}+\frac{x\left(3 x - 2\right)}{3 (x-1)^3} \ln{x}\nonumber\\
F_{8}^{LL}\left[x\right]&=&\frac{x\left(2 + 5 x - x^2\right)}{12 (x-1)^3}-\frac{3 x^2}{6 (x-1)^4}\ln{x}\nonumber\\
F_{8}^{LR}\left[x\right]&=&\frac{1 + x}{2 (x-1)^2}-\frac{x}{(x-1)^3} \ln{x}\nonumber\\
\tilde{F}_{7}^{LL}\left[x\right]&=&\frac{x\left(3 - 5 x\right)}{12 (x-1)^2}+\frac{x\left(3 x - 2\right)}{6
(x-1)^3} \ln{x}\nonumber\\
\tilde{F}_{8}^{LL}\left[x\right]&=&\frac{x\left(3 - x\right)}{4 (x-1)^2}-\frac{x}{2
(x-1)^3}\ln{x}\:.
\end{eqnarray}
In order to make numerical estimates, it is useful to know the values of
these functions evaluated at $x=1$:
\begin{eqnarray}
F_{7}^{LL}\left[1\right]&=&-\frac{5}{72}\:\:,\:\:
F_{8}^{LL}\left[1\right]= -\frac{1}{24}\:\:,\:\:  
F_{7}^{LR}\left[1\right] = \frac{4}{9}\:\:,\nonumber\\  
F_{8}^{LR}\left[1\right]&=& \frac{1}{6}\:\:,\:\: 
\tilde{F}_{7}^{LL}\left[1\right] = -\frac{7}{36}\:\:,\:\: 
\tilde{F}_{8}^{LL}\left[1\right] =-\frac{1}{6}\:.
\end{eqnarray}
 
The two--loop EDMs, on the other hand, depend on the two--loop function
\begin{eqnarray}
F\left[x\right]=\int_{0}^{1} d z\ \frac{z \overline{z}}{x-z \overline{z}}\ \ln{\frac{z \overline{z}}{x}}
\end{eqnarray}
where $\overline{z}=1-z$ with $F\left[1\right]\approx -2.34$.

\end{document}